\documentclass[eqsecnum,amsmath,preprintnumbers,superscriptaddress,nofootinbib,aps,11pt]{revtex4} 

\pdfoutput=1

\usepackage{bm, comment}
\usepackage{MnSymbol}
\usepackage{amsmath}
\usepackage{graphicx}
\usepackage{feynmp}
\usepackage{hyperref}
\usepackage{bbold}
\usepackage{eufrak}
\usepackage{upgreek}
\usepackage{stmaryrd,scalerel}
\usepackage{mathtools}
\usepackage{enumitem}
\usepackage{color} 
\usepackage{etex}

\usepackage{morefloats}

\usepackage{longtable}

\def\beq{\begin{equation}}
\def\eeq{\end{equation}}

\def\bea{\arraycolsep .1em \begin{eqnarray}}
\def\eea{\end{eqnarray}}
\def\Tr{{\rm Tr}}
\def\tr{{\rm tr}}

\def\!!!{\stackrel{!}{=}}

\def\STr{ {\rm STr } }

\def\nn{ \nonumber \\}

\def\eq#1{(\ref{#1})}

\def\s0#1#2{\mbox{\small{$ \frac{#1}{#2} $}}}
\def\0#1#2{\frac{#1}{#2}}

\def\grgl{\:\hbox to -0.2pt{\lower2.5pt\hbox{$\sim$}\hss}{\raise3pt\hbox{$>$}}\:}
\def\klgl{\:\hbox to -0.2pt{\lower2.5pt\hbox{$\sim$}\hss}{\raise3pt\hbox{$<$}}\:}

\begin{document}
\title{Asymptotic Safety within on-shell perturbation theory }
	\author{Kevin Falls}
\address{\footnotesize\mbox{Instituto de Física, Facultad de Ingeniería, Universidad de la República, J.H.y Reissig 565, 11300 Montevideo, Uruguay}}
	\author{Renata Ferrero}
	\address{\footnotesize\mbox{Institute for Quantum Gravity, FAU Erlangen – Nürnberg, Staudtstr. 7, 91058 Erlangen, Germany}}
	\vspace{1pt}
	\date{\today}

\begin{abstract}
We investigate the renormalisation of Einstein gravity using a novel subtraction scheme in dimensional regularisation. The one-loop beta function for Newton’s constant receives contributions from poles in even dimensions and can be mapped to the beta function obtained using a proper-time cutoff.  Field redefinitions are used to remove off-shell contributions to the renormalisation group equations. To check the consistency of our approximations we use a general parametrisation of the metric fluctuation. Within truncations of the derivative expansion and the expansion in Newton’s constant, we show that the parametrisation dependence can be removed order by order. Going to all orders in the scalar curvature an all-order beta function for Newton’s constant is obtained that is independent of the parameterisation. The beta function vanishes at the Reuter fixed point and the critical exponent is in good agreement with non-perturbative calculations.
 Finally, we compare the critical exponent to the counterpart computed via Causal Dynamical Triangulations (CDT).
\end{abstract}

\maketitle

\tableofcontents
\section{Introduction}\label{intro}
The asymptotic safety conjecture \cite{Weinberg:1976xy,Weinberg:1980gg,Parisi:1977uz} proposes that Einstein’s gravity, treated as a quantum field theory (QFT) of the metric tensor, can be ultraviolet complete due to a suitable interacting fixed point of the renormalisation group (RG) \cite{Reuter:1996cp}. 
The availability of non-perturbative renormalisation group (NPRG) methods \cite{Wetterich:1992yh,Morris:1993qb,Ellwanger:1993mw} has allowed for investigations of the so called Reuter fixed point beyond perturbation theory. Consequently, they have become the foundation for the asymptotic safety approach to quantum gravity \cite{Reuter:2019byg, Percacci:2017fkn,Reuter:2001ag,Donoghue:2019clr,Bonanno:2020bil}.

However, the use of NPRG often implies dealing with an effective action that is dependent on unphysical details such as the field parameterisation and the gauge \cite{Falkenberg:1996bq,Nink:2014yya,Falls:2015qga,Gies:2015tca,Ohta:2016npm,Ohta:2016jvw} (see \cite{Falls:2020tmj} for a gauge independent approach that avoids fixing the gauge). These dependencies arise firstly because effective action is off-shell and secondly due to a regulator that breaks diffeomorphism invariance. Consequently disentangling physical information from the flow of the effective action is a particularly arduous task.

To assist in this task, the essential renormalisation group has been introduced in the context of the  NPRG \cite{Baldazzi:2021ydj,Baldazzi:2021orb,Knorr:2022ilz,Knorr:2023usb,Baldazzi:2023pep}. This is a renormalisation scheme which restricts the analysis to the running of the essential couplings \cite{Weinberg:1980gg,Wegner:1974sla}: those couplings which contribute to the scaling of physical observables such as scattering cross sections. In particular, inessential couplings associated with redundant operators are fixed by renormalization conditions achieved by a continuous field reparameterisation along the RG flow \cite{Wegner:1974sla,Pawlowski:2005xe,Latorre:2000qc}.
 The running of the essential couplings should carry physical information, such as scaling exponents at the Reuter fixed point\footnote{See \cite{Kawai:2023rgy} for an alternative approach to fixing an inessential coupling in gravity using a regulator choice.}. While such physical information cannot depend on the choice of gauge or field parameterization, spurious dependencies will occur when non-perturbative approximations are made.

A related approach evaluates the flow of the effective action on-shell and explicitly observes that unphysical dependencies are absent \cite{Benedetti:2011ct,Falls:2014zba,Raul}. Unfortunately, in the context of the NPRG, the flow of the action on-shell still depends on the off-shell effective action \cite{Benedetti:2011ct}. Furthermore, the essential couplings are not exactly those of the on-shell effective action in the presence of the regulator \cite{Baldazzi:2021ydj}. 

Given this situation, investigations based on the NPRG can be complemented with other approaches where the unphysical dependencies can be avoided, even if they fall back on perturbative methods.
One approach is provided by gravity in $d = 2 + \epsilon$ dimensions \cite{Christensen:1978sc,Gastmans:1977ad,Kawai:1989yh,Jack:1990ey,Jack:1990pz,Al-Sarhi:1990nmv,Kawai:1993mb,Aida:1994np,Kitazawa:1995gc,Falls:2017cze,Martini:2021slj,Martini:2021lcx,Martini:2022sll,Martini:2023qkp}, which exhibits an order-$\epsilon $ UV fixed point for Newton’s constant. This approach has motivated the asymptotic safety conjecture since its beginnings \cite{Weinberg:1976xy, Weinberg:1980gg} and served as the initial testing ground for asymptotic safety. An important lesson that one learns from this approach is that the dependence on the gauge and parameterization is absent at one-loop once field redefinitions are used to fix the renormalization of an inessential coupling \cite{Kawai:1989yh,Falls:2017cze}. For example, if one uses a redefinition in pure gravity to ensure the cosmological constant is not renormalised then the beta function for the dimensionless Newton's coupling $\tilde{G}$ (in units of the RG scale) is given by
\beq \label{beta_intro}
\beta_{\tilde{G}} = \epsilon \, \tilde{G} - \frac{38}{3} \tilde{G}^2 \,, 
\eeq
independent of the gauge and the parameterisation.
The main limitation of working in $d = 2 + \epsilon$ dimensions is that extending it to $d=4$ means taking $\epsilon=2$, which seems beyond the validity of perturbation theory.  Moreover, pure two-dimensional gravity is topological, as the integrated curvature scalar is also the Euler’s density. This leads to the problem of kinematical poles \cite{Kawai:1989yh,Jack:1990ey} which seem to prevent a meaningful extension of the scheme beyond one-loop.

The purpose of this work is to use perturbative methods in four dimensions to investigate asymptotic safety.  Investigations of this type have been put forward in
\cite{Niedermaier:2009zz,Niedermaier:2010zz}.
At least at one-loop one can use a proper-time regulator  which preserves gauge symmetry - we refer the reader to \cite{Bonanno:2000yp,Mazza:2001bp,Bonanno:2004sy,deAlwis:2017ysy,Bonanno:2019ukb,Abel:2023ieo,Giacometti:2024qva,Glaviano:2024hie} for investigations with an RG improved proper-time flow equation and its connection to the Wilsonian RG.  Alternatively, one can use a generalised version of dimensional regularisation, which does not single out any particular dimension when subtracting poles as $d\to d_c$. In fact, the idea of subtracting poles in more dimensions was initiated by  \cite{Jack:1990pz,Al-Sarhi:1990nmv}. Furthermore, this approach has been advocated by Weinberg \cite{Weinberg:1980gg} but only applied to gravity recently \cite{Kluth:2024lar}.
In this approach, instead of fixing one critical dimension $d_c$ to determine the counter terms, one can sum over all $d_c$ so that all poles are removed.
At one-loop poles appear in all even dimensions $d_c$ beyond one-loop poles appear in new dimensions including fractional dimensions.
Here we use such a scheme combined with the non-minimal subtraction scheme proposed in \cite{Martini:2021slj} which treats differently the dependence on $d$  which originates from components of the metric tensor $g_{\mu\nu}$.
In particular $g^{\mu}_\mu=d$ is kept fixed in the counter term as $d \to d_c$ such that a finite subtraction is made.

Both proper-time regularisation and dimensional regularisation approaches allow for fully functional approximations keeping classes of invariants to all orders.
In this way, the RG improvement of the one-loop flow equation appears similar to the NPRG, even if it is not exact \cite{Litim:2001ky}.   
Combined with the newly developed essential RG we find suitable approximations where the unphysical dependencies are under control, including keeping all orders in the scalar curvature.

Our main result is to obtain a value for the relevant critical exponent at the Reuter fixed point, which is explicitly independent of the parametrisation. Although we do not test it explicitly, we also expect our result to be gauge-independent by standard arguments.
This can be achieved since, within our one-loop approximation, the essential scheme allows us to fix all off-shell terms in the effective action, and thus the physical running is obtained from an on-shell projection. 
This gives a beta which extends \eq{beta_intro} both away from two dimensions and to higher orders in $\tilde{G}$.
Remarkably we also find that dimensional regularisation and the proper-time regularisation are in exact agreement at one-loop. The value of the exponent can also be compared with those obtained from the NPRG \cite{Baldazzi:2021orb,Knorr:2023usb,Baldazzi:2023pep} and from recent lattice studies \cite{Ambjorn:2024qoe} in the form of Causal Dynamical Triangulations (CDT) \cite{Ambjorn:2020rcn,Ambjorn:2012jv,Loll:2019rdj}.

\bigskip

This paper is structured as follows. In section \ref{sec:oneloop} we derive the one-loop effective action for Einstein gravity and in section \ref{sec:MES} we review the minimal essential scheme. In section \ref{sec:dim_reg} we use dimensional regularisation to compute the traces of the effective action and project onto the on-shell effective action. 
This calculation involves subtracting poles in zero, two and for dimensions and leads to a beta function that is third order in Newton's constant $G$.
In section \ref{sec:essential} we generalise the idea of the essential RG in perturbation theory, by means of a proper-time cutoff, and  we introduce the minimal essential scheme (MES) in perturbation theory, making manifest the relation to the on-shell projection. Furthermore, we apply this scheme and derive the beta functions to order curvature square. Here we investigate the parameterization dependence of the beta function for $G$ and note that it is absent if we expand to order $G^3$. The map to dimensional regularisation is then found in this approximation. Further, consistent approximations are described that include progressively higher order operators and higher orders in $G$.
Following this idea to higher orders, section \ref{sec:allcurvature} extends our approximations to account for all orders in the scalar curvature $R$ in the proper-time flow equation. The results in this section are the main outcome of this work.
We obtain the beta functions and the critical exponent to all curvature order on a maximally symmetric background which are independent of the parameterization as they are obtained on-shell.
The equivalence between the proper-time and the dimensional regularisation is shown in section \ref{sec:equivalence}. In section \ref{sec:crit_exp} we analyse the connection between the essential RG scheme and lattice computations and we conjecture the connection with the critical exponent recently computed in CDT. Finally, in section \ref{sec:conclusions} we conclude and give an outlook.

For the convenience of the reader, the appendices collect the technology needed to evaluate one-loop traces using results that can be found in the literature \cite{DeWitt:1960fc,Barvinsky:1985an,Barvinsky:1990up,Barvinsky:2021ijq,Barvinsky:2024kgt,Benedetti:2010nr,Groh:2011dw,Kluth:2019vkg,Ferrero:2023xsf,Benedetti:2014gja,Falls:2016msz,Banerjee:2023ztr}.
Appendix \ref{app_trace} contains the computation of the one-loop traces for the linear parametrisation. In appendix \ref{app:noncomm} we generalise the computation to a general parametrisation evaluating non-commuting traces. 
Appendix \ref{app:off} contains the off-diagonal heat kernel coefficients \cite{Groh:2011dw} which we use in our calculation,  appendix \ref{app:sphere} summarises the derivation of the coefficients on a sphere to all order, while in appendix~\ref{app:hyperboloid} the spectral sums for spheres and hyperboloids are detailed.

\section{One-loop effective action}\label{sec:oneloop}

We start considering Einstein-Hilbert gravity in  Euclidean signature. The background effective action at one-loop reads
\beq 
\Gamma = S + \frac{1}{2} \Tr \log \left[ K^{-1} (S^{(2)} + S^{(2)}_{\rm gf})\right] -  \Tr \log[ \mathcal{Q}_{\rm FP}]\,.
\label{1-loop}
\eeq
All our results will be derived from the renormalisation of this effective action which is evaluated by taking the background field $\bar{g}_{\mu\nu}$ to equal the mean field  $\bar{g}_{\mu\nu} = g_{\mu\nu}$, as per the 
background field method.  
In \eq{1-loop} $S$ is the Einstein-Hilbert action, with the addition of the topological term needed to renormalise the theory at one-loop in $d=4$ dimensions. The other curvature squared terms can be removed by field redefinitions since they, i.e., by using the equations of motion.  Thus, the action $S$ reads 
\begin{equation}\label{eq:eh}
S = \int \text{d}^d x \; \sqrt{g} \left(\frac{\rho}{8\pi}-\frac{R}{16\pi G} +  \vartheta \mathfrak{E}(g)  \right) \,,	
\end{equation}
where 
\beq
\mathfrak{E} = R^2 - 4 R_{\mu \nu} R^{\mu \nu} + R_{\rho \sigma \mu \nu} R^{\rho \sigma \mu \nu}\,,
\eeq
is a combination of curvature-squared terms, $G$ is Newton's constant, $\rho$ is the cosmological constant and $\vartheta$ is the coupling to $\mathfrak{E}$. In the more standard convention the cosmological constant is usually defined as $\Lambda_{\rm cc} = G \rho$.
Since later we will fix the value of $\rho$ we opt mostly for the unconventional notation. 
$S^{(2)}$ is the hessian of the action,  $S^{(2)}_{\rm gf}$ is the hessian of a gauge fixing action, $K^{-1}$ is the inverse of the ultra local DeWitt metric $K$, which appears as the tensorial structure in front of the Laplacian. $K$ and its inverse are
 \beq\label{eq:dewitt}
 K^{\mu\nu, \rho \lambda}(x,y) = \frac{1}{64 \pi G} \sqrt{ g} (  g^{\mu\rho}  g^{\nu\lambda}+   g^{\mu\lambda}  g^{\nu\rho} -   g^{\mu\nu}  g^{\rho\lambda}) \delta(x-y) \,,
 \eeq
\beq
(K^{-1})^{\mu\nu, \rho \lambda}(x,y) = \frac{64 \pi G} {\sqrt{ g}} \left(  g^{\mu\rho}  g^{\nu\lambda}+   g^{\mu\lambda}  g^{\nu\rho} -  \frac{2}{d-2} g^{\mu\nu}  g^{\rho\lambda}\right) \delta(x-y) \,.
\eeq
In \eqref{1-loop} $ \mathcal{Q}_{\rm FP}$ is the ghosts operator. 
 For the gauge fixing action we take 
	\begin{equation}
		S_{{\rm gf} }[g;\bar{g}] = \frac{1}{2 } \int {\rm d}^d x \sqrt{\bar{g}} F^{\nu}  \bar{g}_{\mu\nu} F^{\mu} \,,
	\end{equation}
 where $\bar{g}_{\mu\nu}$ is the background field, and we use the background covariant harmonic gauge
	\begin{eqnarray}
		F^\mu
		&=& \frac{1}{\sqrt{16 \pi G}}  \left( \bar{g}^{\mu \lambda} \bar{g}^{ \nu \rho}  - \frac{1}{2} \bar{g}^{\nu\mu}   \bar{g}^{\rho \lambda}     \right)  \bar{\nabla}_\nu  g_{\lambda \rho}\, .\label{eq:gauge}
	\end{eqnarray}
This leads to the ghost operator
	\begin{equation}
		(\mathcal{Q}_{\text{ FP}})^{\mu}\,_{\nu} c^{\nu} \equiv - \sqrt{16 \pi G }\mathcal{L}_{c} F^{\mu} =  - \left(  \bar{g}^{\mu \lambda} \bar{g}^{ \nu \rho}  - \frac{1}{2}\bar{g}^{\nu\mu}   \bar{g}^{\rho \lambda}     \right)  \bar{\nabla}_\nu ( g_{ \rho \sigma} \nabla_{\lambda} c^{\sigma}  + g_{ \lambda \sigma} \nabla_{\rho} c^{\sigma}) \,,
	\end{equation}
	which enters the ghost action
	\begin{equation}
		S_{\rm{gh}}[g, c ,\bar{c};\bar{g}] = \int {\rm d}^d x\sqrt{\bar g}   \bar{c}_\mu (\mathcal{Q}_{\text{ FP}})^{\mu}\,_{\nu} c^{\nu}  \,.
	\end{equation}

The effective action \eq{1-loop} depends on how we parameterise quantum fluctuations of the metric. Here, to demonstrate that physical quantities are independent of this choice, we wish to consider an arbitrary parameterisation of the quantum metric $\hat{g}_{\mu\nu}= \hat{g}_{\mu\nu}(\hat{\phi}, \bar{g})$ in terms of a tensor field $\hat{\phi}_{\mu\nu}$ and a background metric field $\bar{g}_{\mu\nu}$. To employ the background field method we write $\hat{\phi}_{\mu\nu}= \bar{g}_{\mu\nu} + \hat{h}_{\mu\nu}$ where $\hat{h}_{\mu\nu}$ is a flucation field which is coulped to a source in the path integral. 
In particular, expanding up to the second order in $\hat{h}_{\mu\nu}$ we consider parameterisations $\hat{g}_{\mu\nu}(\hat{\phi}, \bar{g})$ such that \cite{Gies:2015tca}
\begin{equation}\label{eq:expand}
    \hat{g}_{\mu \nu} = \bar{g}_{\mu \nu} +\hat{h}_{\mu \nu}+ \frac{1}{2}\left( \tau_1 \hat{h}_{\mu \rho} \hat{h}^{\rho}_{\nu}+\tau_2 \hat{h} \hat{h}_{\mu \nu}+ \tau_3 \bar g_{\mu \nu} \hat{h}_{\rho \sigma}\hat{h}^{\rho \sigma}+ \tau_4\bar g_{\mu \nu}\hat{h}^2\right) + O(\hat{h}^3)\,,
\end{equation}
 where indices are raised and lowered with the background field $\bar{g}_{\mu\nu}$ in this expression and $\hat{h} = \bar{g}^{\mu\nu} \hat{h}_{\mu\nu}$.
 Therefore, physics should not depend on the parameters $\tau_i$. 
 
 To obtain $\Gamma$ we couple the source term to the quantum fluctuations $\hat{h}_{\mu\nu}$. In this way the hessian of the action which enters \eq{1-loop} is given by 
\beq
S^{(2) \mu\nu,\rho \lambda}(x,y) = \frac{\delta S}{ \delta h_{\mu\nu}(x) \delta h_{\rho  \lambda}(y) }\,,
\eeq
where $h_{\mu\nu}$ is a fluctuation field,  i.e., the expectation value of a quantum fluctuation $h_{\mu\nu} = \langle \hat{h}_{\mu\nu}\rangle$. The effective action $\Gamma= \Gamma[h;\bar{g}]$ depends on the background field $\bar{g}_{\mu\nu}$ and the fluctuation field $h_{\mu\nu}$.
As we have said, following the background field method, we put $h_{\mu\nu} =0$ in \eq{1-loop} and hence the action is only a functional of the background field with $\bar{g}_{\mu\nu} = \langle \hat{\phi}_{\mu\nu} \rangle$. The equivalence theorem and background field method guarantee that amplitudes can be computed from this functional.  
 By the background field method, we can compute amplitudes of $\hat{\phi}_{\mu\nu}$ from correlation functions obtained from the background field  derivatives of $\Gamma$ at $h_{\mu\nu}=0$ \cite{Abbott:1981ke}. Additionally, the equivalence theorem \cite{Chisholm:1961tha,Kamefuchi:1961sb,Efimov:1972juh,Kallosh:1972ap,Arzt:1993gz,Tyutin:2000ht} states that amplitudes are independent of the parameterisation, i.e., independent of $\hat{\phi}_{\mu\nu}(\hat{g},\bar{g})$ and hence the values of the parameters $\tau_i$.

Thus, from now on, we can work with only one metric, which we denote by $g_{\mu\nu}$, to lighten the notation. $\Gamma$ is therefore a functional of a single metric,  $\Gamma[g] = \Gamma[0;g]$. Using the equations of motion for $S$ in the rhs all terms that depend on the choice of the  parameters $\tau_i$ vanish. This is easily seen by noting the parts of the hessian proportional to  $\tau_i$ vanish on the equations of motion. One can also show the same is true of the gauge dependence \cite{Dou:1997fg,Benedetti:2011ct,Falls:2017cze}, although we will not keep track of the gauge dependence here.

Let us note that in $d=2$, $d=3$ and $d=4$ the term $\int \text{d}^dx \sqrt{g} \mathfrak{E}(g)$ is either zero or an invariant and thus it does not contribute to the functional derivatives of the action. Since we are mostly interested in the case $d=4$ we will neglect the terms coming from the functional derivatives of  $\int \text{d}^dx \sqrt{g} \mathfrak{E}(g)$ in non-integer dimensions to obtain expressions for general $d$ between $d=2$ and $d=4$.  
Thus, in the equation of motion and the hessian we can neglect terms proportional to $\vartheta$. This simplifies our calculations. As a result, our interpolations are generally unreliable for non-integer dimensions and $d>4$; they are valid only for $d=2$, $d=3$, and $d=4$.

In this work we will use various approximations to obtain the renormalisation group flow of couplings in quantum gravity. Firstly, the divergences originate from the one-loop effective action, hence the scheme is a one-loop improved type and therefore misses information starting at two-loops. Secondly, we make truncations of the effective action to include only certain terms (e.g. expansions in curvature invariants). Thirdly, we will resort to expansions in Newton's coupling $G$. A combination of the last two expansions counts $G\rho$ as the same order as curvature, as suggested by its dimensionality (e.g. $R G \rho$ is the same order as $R^2$).

 \section{The minimal essential scheme in quantum gravity}\label{sec:MES}
In this section, we elaborate on the schemes introduced in \cite{Baldazzi:2021ydj,Baldazzi:2021orb} which were introduced to utilise reparameterisation invariance in QFT.
This arises because we have the freedom to perform field redefinitions
\beq
\phi \to F[\phi']
\eeq
in the path integral.
Two QFTs that are related by a field redefinition will produce the same scattering amplitudes (at least in perturbation theory) moreover there will exist a dictionary between observables with correlation functions of $O(x)$ being equal to  $O'(x)$ where
\beq
O'[\phi'] = O[F[\phi']]\,.
\eeq

If the value of a coupling constant can be changed by making a field redefinition then this coupling is inessential. It follows that by making a continuous field redefinition along the RG flow we can fix the values of the inessential couplings. 

In principle, we can fix the inessential couplings to any value we choose. In practice, however, the values of observables computed in an approximation scheme will depend on this choice due to errors arising from the approximation, in particular, the truncation of the action to include only certain terms, i.e., the curvature up to some fixed order. On the other hand, we expect these errors will vanish at the Gaussian fixed point which is the free theory (reached in the IR in the case of gravity). At the Gaussian fixed point of a scalar field theory the inessential couplings are those which multiply operators that vanish on the classical equation of motion of the scalar, i.e., when $\partial^2 \phi =0$.

The MES fixes the values of the inessential couplings to the values they take at the Gaussian fixed point. 
The rational for this scheme is threefold:
\begin{itemize}

\item First, at the Gaussian fixed point the inessential couplings are easy to identify and thus the scheme is straightforward to implement. For the case of gravity the Gaussian fixed point is where $G=0$ and $\Lambda_{\rm cc}=0$. The inessential couplings are the couplings to operators which vanish when the vacuum Einstein equations 
\beq
R_{\mu\nu} = 0
\eeq
hold, apart from the Einstein-Hilbert term.
These couplings (e.g. the coefficients of $R^2$ and $R_{\mu\nu}R^{\mu\nu}$) vanish at the Gaussian fixed point so they remain zero in the MES.
In addition we fix the dimensionless cosmological constant to its value at the Gaussian fixed point 
\beq
 \mu^{-d} \rho = \tilde{\rho}_{\rm GFP}
\eeq
where $\mu$ is the RG scale. Generally, $\tilde{\rho}_{\rm GFP}$ is non-zero\footnote{An exception is the scheme used in \cite{Kluth:2024lar} where $\tilde{\rho}_{\rm GFP} = 0$.} and its value depends on the regularisation and subtraction scheme.

\item Secondly, this scheme reduces the complexity of computations since terms in the action which would typically arise and contribute to the flow equations are set to zero.

\item Finally, since in the MES the inessential couplings do not flow away from their values at the Gaussian fixed point, we expect to minimise errors.   

\end{itemize}

Even though it is important to explore different schemes and examine their dependence, in this paper, we will adopt the MES for the reasons outlined above.

\section{Dimensional regularisation}\label{sec:dim_reg}
In this section we use a variant of dimensional regularisation that generalises the scheme of \cite{Martini:2021slj,Martini:2022sll} where divergences that appear as  $d \to 2$ were investigated. As suggested in \cite{Weinberg:1980gg} it is straightforward to keep also divergences that appear in other dimensions. So here we keep one-loop divergences that also appear as $d\to 0$ and $d\to 4$. In this manner, we have a form of dimensional regularisation that keeps power-law divergences in addition to logarithmic ones in $d=4$ dimensions. To ensure that our results are physical we will use the MES scheme for gravity introduced in \cite{Baldazzi:2021orb}. Here this scheme dictates exactly how the equations of motion are used to obtain ``essential'' beta functions. At one-loop the essential couplings that get renormalised (in particular in $d=4$ dimensions) are $\vartheta$ and a dimensionless product $\eta \propto G \rho^{\frac{d-2}{d}}$.

We now want to find the form of the UV singularities that occur for $d\leq 4$. To keep all types of UV singularities we exploit dimensional regularisation and remove poles at $d=0$, $d=2$, and $d=4$. In this manner, we do not neglect the singularities responsible for asymptotic safety close to two dimensions.
We could continue to add more terms by including singularities when $d \to d_c$ for all even integers.  For each $d_c$ we will have terms where $2 n_\partial+ n_{\rm cc} = d_c$ where $n_\partial \geq 0$ is the number of derivatives that act on the metric in the term and $n_{\rm cc} \geq 0$ is the number of powers of the cosmological constant $\Lambda_{\rm cc} = G \rho$. 
So, for example, for $d_c =0$ we have only the term where $n_\partial = 0$ and $n_{\rm cc} =0$.

Including the singular terms for $d_c\leq 4$ entails that we expand the trace to the second order in curvature and treats the cosmological constant $\Lambda_{\rm cc} = G \rho$ as a power of curvature. We can also neglect covariant derivatives of the curvature since these give only a boundary term which we ignore. Thus the singular part of the effective action we are interested in is given by
\beq
\Gamma = \int \text{d}^d x \; \sqrt{g} \left( \frac{a_0(d)}{d} \mu^d  + \frac{a_2(d)}{d-2}\mu^{d-2} R + \frac{a_{4}(d)}{d-4} \mu^{d-4}  G \rho R   + \frac{a_{4}'(d)}{d-4} \mu^{d-4} \mathfrak{E} \right) + {\rm finite \, terms}\;,
\label{Gamma_sing}
\eeq
where we have exploited the equations of motion
\beq
G \rho g^{\mu\nu} + R^{\mu\nu}-\frac{1}{2}Rg^{\mu \nu} =0 \,
\eeq
to keep only a minimal set of terms. Specifically, the terms $R^2$, $R_{\mu\nu} R^{\mu\nu}$ and $(G \rho)^2$ can be replaced by terms we have kept through a field redefinition which produces terms proportional to the equations of motion. The minimal set of singular terms that we have chosen ensures that we do not need to introduce counter terms outside the Einstein-Hilbert action apart from the topological term. Furthermore, the vacuum energy is only renormalised due to the singular term in $d=0$ dimensions. This follows the MES for gravity where we fix inessential couplings to the values they take at the Gaussian fixed point (GFP) \cite{Baldazzi:2021orb}.

The traces are computed using heat kernel techniques \cite{Barvinsky:1985an,Barvinsky:1990up,Barvinsky:2021ijq,Barvinsky:2024kgt,Benedetti:2010nr,Groh:2011dw} as detailed in the appendix~\ref{app_trace} for the linear parameterisation ($\tau_i=0$). Furthermore, for a general parametrisation \eqref{eq:expand} traces of non-commuting operators have to be evaluated which are carried out in appendix~\ref{app:noncomm}.\footnote{Note that the off-diagonal heat kernel machinery \cite{Groh:2011dw} enables the computation of traces of non-minimal operators. This setup is somehow complementary and orthogonal to the commutator-based method of \cite{Barvinsky:2021ijq,Barvinsky:2024kgt}. } This lets us check that the parameterisation dependence cancels when we use the equations of motion.

The local part  coming from the traces in $\Gamma$ is UV divergent and expressed as an integral over proper-time $s$. We extract the UV singular parts by noting that integrals of the form $\int_0^\infty \text{d}s s^{\frac{d_c-d}{2} -1}$ have the UV singular part 
\beq
\int_0^\infty \text{d}s s^{\frac{d_c-d}{2} -1} \sim  - 2 \frac{\mu^{d-d_c}}{d-d_c}
\eeq
as $d \to d_c$, where $\mu$ is the usual mass scale introduced in dimensional regularisation. One then finds the following coefficients 
\bea
a_0(d) &=& \frac{1}{(4 \pi)^{\frac{d}{2}} }  \frac{1}{2} (d-3) d\\
a_2(d) &=& \frac{1}{(4 \pi)^{\frac{d}{2}} } \frac{1}{12} \left(d^2-3 d-36\right)\\
a_4(d) &=&  \frac{1}{(4 \pi)^{\frac{d}{2}} }    \frac{d^3+19 d^2-566 d+1200}{120 (d-2)}\\
a_4'(d)&=& \frac{1}{(4 \pi)^{\frac{d}{2}} } \frac{1}{360} \left(d^2-33 d+540\right)\,.
\eea
We note that since these expressions have been obtained using the equations of motion they are in principle independent of the gauge in addition to the choice of field variables \eqref{eq:expand}.
We see this explicitly since the coefficients are independent of the parameters $\tau_i$.
 Each $a_{d_c}(d)$ is a function of the dimension. However,
\beq
\frac{a_{d_c}(d)}{d- d_c} =  \frac{a_{d_c}(d_c)}{d- d_c} + {\rm finite \, terms}\;,
\eeq
thus a standard minimal subtraction would set the counterterms to 
\beq
S_{\rm ct} =- \int \text{d}^d x \; \sqrt{g} \left( \frac{a_0(0)}{d} \mu^d  + \frac{a_2(2)}{d-2}\mu^{d-2} R + \frac{a_{4}(4)}{d-4} \mu^{d-4}  G \rho R   + \frac{a_{4}'(4)}{d-4} \mu^{d-4} \mathfrak{E} \right)\;.
\eeq
 In \cite{Martini:2021slj} it has been suggested to keep $g^{\mu}\,_\mu \equiv d $ distinct from $d_c$ when making the subtraction since otherwise one is identifying the number of components of the metric with the regularisation parameter. Making this distinction amounts to keeping the rational dependence on $d$ while sending $1/(4\pi)^{d/2} \to1/(4\pi)^{d_c/2}$. Thus, the coefficients in the counter term are given by
\bea
\label{abars}
\bar{a}_0(d) &=&   \frac{1}{2} (d-3) d\nn
\bar{a}_2(d) &=& \frac{1}{(4 \pi)} \frac{1}{12} \left(d^2-3 d-36\right)\nn
\bar{a}_4(d) &=&  \frac{1}{(4 \pi)^{2} }   \frac{d^3+19 d^2-566 d+1200}{120 (d-2)}\nn
\bar{a}_4'(d)&=& \frac{1}{(4 \pi)^{2}} \frac{1}{360} \left(d^2-33 d+540\right)\,,
\eea
such that
\beq
S_{\rm ct} =- \int \text{d}^d x \; \sqrt{g} \left( \bar{a}_0(d) \frac{\mu^d}{d}  + \bar{a}_2(d) \frac{\mu^{d-2}}{d-2} R + \bar{a}_{4}(d) \frac{\mu^{d-4}}{d-4}  G \rho R   + \bar{a}_{4}'(d) \frac{\mu^{d-4}}{d-4} \mathfrak{E} \right) \;.
\label{counter_terms}
\eeq
The value of $\bar{a}_2$ was obtained in \cite{Martini:2021slj} (there denoted $-A$), where only the divergences for $d_c=2$ were kept and agrees with \cite{Bastianelli:2013tsa} for $d=4$. Here we have extended the scheme by adding the counter terms for $d_c= 0$ and $d_c =4$.
The first term in \eq{counter_terms} renormalises the vacuum energy and counts precisely the number of polarisations of the graviton in $d$ dimensions.
The second and third terms in \eq{counter_terms} then renormalise Newton's constant $G$ while the last divergence is proportional to the topological term. We note that $\bar{a}_4(d)$ and $\bar{a}'_4(d)$ generalise the results of \cite{Gibbons:1978ac,Christensen:1979iy} to $d$ dimensions and are in agreement with \cite{Bastianelli:2022pqq} in arbitrary $d$.

Let us note that if we had not sent $d \to d_c$ in the factors $1/(4\pi)^{d/2}$ (i.e., if replace $\bar{a}(d)$ with $a(d)$ in the counter term), the critical exponent computed later in this section would be unaffected. This observation, in fact, motivates us to use a proper-time cutoff in section~\ref{sec:essential}.
As will be further discussed there, and in section~\ref{sec:equivalence}, we will see that the two schemes are related by a constant rescaling of the coupling constants.

\subsection{Fixing the inessential coupling}
As we have said, we are going to fix the inessential couplings to the values they take at the GFP.
The GFP is found by expanding the metric around flat spacetime and sending $G \to0$ such that we have a free theory.
Thus for the free theory we simply have the counter term 
\beq
S_{\rm ct} =- \int \text{d}^d x \; \sqrt{g}  \bar{a}_0(d) \frac{\mu^d}{d}  \;.
\label{counter_terms_GFP}
\eeq
Defining the dimensionless couplings $\tilde{G} = \mu^{d-2} G$ and $\tilde{\rho} = \mu^{-d} \rho$  the beta functions for the free theory are 
\beq \label{beta_rho_dim_reg}
\beta_{\tilde{\rho}} = - d \tilde{\rho} + 8\pi  \bar{a}_0  
\eeq 
and 
\beq
\beta_{\tilde{G}} = (d-2) \tilde{G} \,.
\eeq
Thus at the GFP we have
\beq \label{rho_con_dim_reg}
\frac{\tilde{\rho}}{8\pi} = \frac{\bar{a}_0}{d} =  \frac{1}{2} (d-3) \,.
\eeq
The MES adopts \eq{rho_con_dim_reg} as a renormalisation condition such that the dimensionless inessential coupling does not flow away from its value at the GFP. 

\subsection{Essential beta function}
From \eq{counter_terms} with \eq{abars} one can evaluate the flow equation and compute the beta functions. In fact by requiring that 
\begin{equation}
    \mu \frac{d}{d\mu} \Gamma = 0 \qquad \to \qquad  \mu \frac{d}{d\mu} S = - \mu \frac{d}{d\mu} S_\text{ct}\,.
\end{equation}
Note that because we do not further expand  $S_\text{ct}$  in $\hbar$,  i.e., by expanding the couplings, this corresponds to a one-loop improved RG flow. 

The beta function for Newton's constant is given by
\beq
\beta_{\tilde{G}} = (d-2) \tilde{G} +\frac{1}{3} ((d-3) d-36) \tilde{G}^2 +  \frac{(d (d (d+19)-566)+1200) \tilde{G}^3 \tilde{\rho} }{120 \pi  (d-2)} \, .
\eeq
With the condition \eq{rho_con_dim_reg} we have 
\beq
\beta_{\tilde{G}} = (d-2) \tilde{G} +\frac{1}{3} ((d-3) d-36) \tilde{G}^2 +  \frac{(d-3)(d (d (d+19)-566)+1200) \tilde{G}^3 }{30  (d-2)}\,.  
\eeq
 The coefficient of the $\tilde{G}^2$ was first obtained in \cite{Falls:2015qga} using a covariant momentum cutoff and reproduced in \cite{Martini:2021slj} in dimensional regularisation. As $d\to 2$ the coefficient agrees with \eq{beta_intro}. 
 The third term proportional to $\tilde{G}^3$ results from removing poles as $d\to 4$. Later we will extend this beta function to all orders in $\tilde{G}$. We note that the $\tilde{G}^3$ coefficient is singular as $d\to 2$, this is an example of a kinematical pole due to the topological nature of gravity in two dimensions.
 
Let us define the  ``essential'' coupling
\beq \label{eta_def}
\eta \equiv G \left(\frac{\rho}{4 \pi (d-3)}\right)^{\frac{d-2}{d}}\,,
\eeq
which is dimensionless and hence invariant under a rescaling of the metric.
Then the condition \eq{rho_con_dim_reg} implies that 
\beq \label{eta_MES_DR}
\eta = \tilde{G} \,,
\eeq
and we obtain the 
\beq \label{betaeta}
\beta_\eta =  (d-2) \eta +\frac{1}{3} ((d-3) d-36) \eta ^2 +  \frac{(d-3)(d (d (d+19)-566)+1200) \eta^3 }{30  (d-2)} \,.
\eeq
Notice that this is just one choice of $\eta$, in order to find a dimensionless combination of $G$ and $\rho$. It is chosen to realise \eq{eta_MES_DR}. Although the $d-3$ pole might seem worrisome, it cancels because $\tilde{\rho}$ vanishes when $d=3$, resulting in a combination which is finite and non-zero even in three dimensions. These elements motivate our choice. 

For $d=4$ this beta function has a fixed point at $\eta_\star = \frac{1}{87} \left(\sqrt{2905}-40\right)\approx 0.16$ where the critical exponent is given by
\beq \label{theta_order_2}
\theta = -\left. \frac{\partial \beta_\eta }{\partial \eta}\right|_{\eta = \eta_\star} =  \frac{4}{261} \left(581-8 \sqrt{2905}\right) \approx 2.296\,.
\eeq
This value is numerically in good agreement with calculations using the functional renormalisation group in various approximations which also use the MES \cite{Baldazzi:2021orb,Knorr:2023usb,Baldazzi:2023pep}.
Let us stress that this result should be independent of unphysical details such as the parameterisation of the metric or the gauge since we have used the equations of motion. On the other hand, it does depend on our choice to fix the dimensionless cosmological constant $\tilde{\rho}$ to its value at the GFP. 
This is related to the fact that one can only measure dimensionless couplings, and thus to have a physically meaningful exponent, we have one coupling fixed in units of the cutoff. If we had included matter, we might have chosen to fix, for example, the mass of a particular particle. For Einstein-Hilbert gravity, the cosmological constant is a unique choice other than fixing Newton's constant itself. We note that this is not necessarily the same as keeping the volume operator fixed in the path integral. 

\begin{figure}
		\centering
  	\includegraphics[scale=0.29]{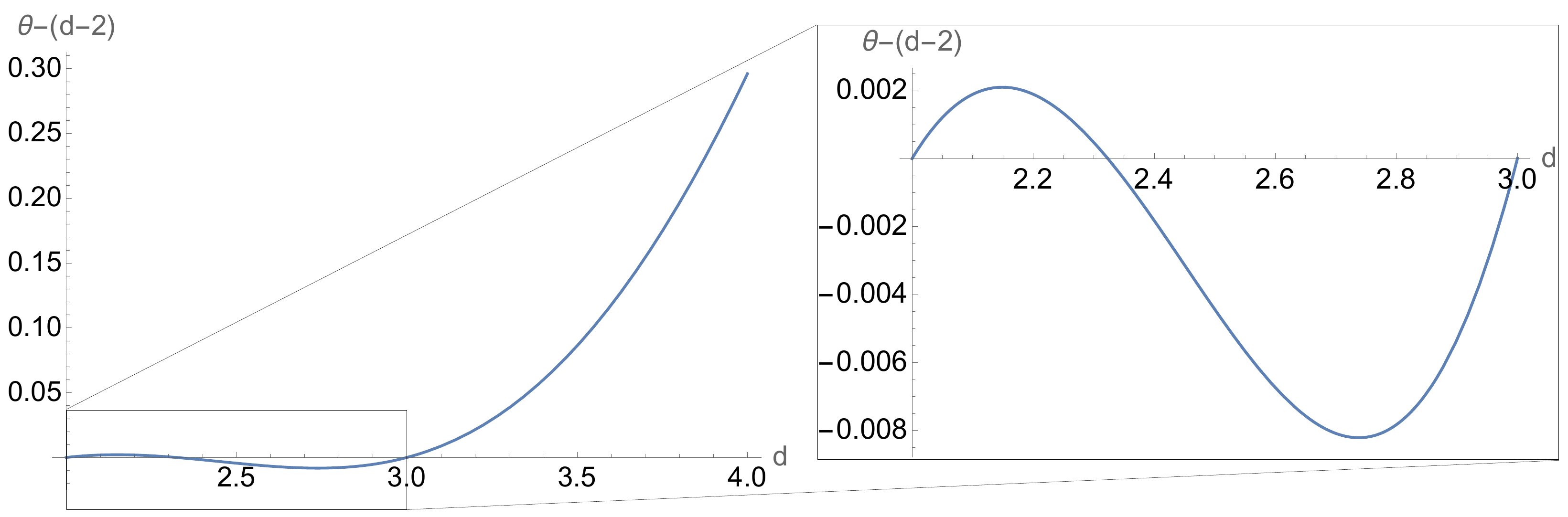}\quad 
		\caption{Plot of the critical exponent as a function of the dimension.}\label{fig:theta}
	\end{figure}

There is also a fixed point in all $2<d<4$. We can characterize quantum corrections to the scaling by $\theta-(d-2)$, which vanishes both as $d\to2$ and in $d=3$ dimensions where the coefficient of the $\eta^3$ term vanishes. The value of  $\theta-(d-2)$ between two and four dimensions is shown in Fig.~\ref{fig:theta}. We observe the quantum corrections remain very small between two and three dimensions and grow more rapidly for $d>3$.

The running of the coupling to the topological operator is given by
\beq \label{Euler_running_DR}
\mu \partial_\mu \vartheta_\mu =  \mu^{d-4} \frac{1}{(4 \pi)^{2}} \frac{1}{360} \left(d^2-33 d+540\right) 
\eeq
which runs logarithmically in $d=4$ dimensions.

\section{Essential RG with a proper-time cutoff}\label{sec:essential}
Having computed the essential beta function using dimensional regularisation, we now want to use a proper-time regulator which at one-loop allows for an alternative gauge invariant regularisation of the theory.
As we have anticipated, this approach will lead to results which are in exact agreement with dimensional regularisation, provided the corresponding approximations are used.

\subsection{Generalised proper-time flow equation}
As introduced in section \ref{sec:oneloop}, at one loop the effective action is given by 
\beq
\Gamma[\phi] = S[\phi] + \frac{\hbar}{2} \Tr \log K^{-1}[\phi] S^{(2)}[\phi]/M^2 + O(\hbar^2)\;.
\eeq
Here we have suppressed the ghosts and denote the field $\phi$ for a general theory. We have included $M^2$ which is an arbitrary scale. This scale appears for dimensional reasons and we can identify with the UV cutoff hence we will take the limit $M \to \infty$ in the end. 
Expressing the trace term with a proper-time integral we have 
\beq
\Gamma[\phi] = S[\phi]  - \frac{\hbar}{2} \int^{\infty}_0 ds \frac{1}{s} \Tr \left( \exp( -s  K^{-1}[\phi] S^{(2)}[\phi]) -  \exp( -s M^2) \right) + O(\hbar^2)\;,
\eeq
then cutting off the upper limit at $1/k^2$ the lower limit at $1/M^2$ we get an infrared regulated and UV regulated effective action  
\beq
\Gamma_{k,M}[\phi] = S[\phi]  -  \frac{\hbar}{2} \int^{k^{-2}}_{M^{-2}} ds \frac{1}{s} \Tr \left( \exp( -s  K^{-1}[\phi] S^{(2)}[\phi]) -  \exp( -s M^2) \right) + O(\hbar^2)\;.
\eeq
Taking a $k$ derivative and taking the limit $M \to \infty$ we get the flow equation at order $\hbar$
\beq
 k \partial_k  \Gamma_k[\phi] = \hbar \;\Tr \, \exp(- K^{-1} S^{(2)}_k[\phi] k^{-2}) +    O(\hbar^2)\;.
\eeq    
The proper-time flow equation is a one-loop RG-improved flow equation given by replacing the classical action and the metric by the effective action and an RG-improved DeWitt metric $K_k$ \eqref{eq:dewitt}:   
\beq \label{Proper-time}
 k \partial_k  \Gamma_k[\phi] = \Tr \, \exp(- K^{-1}_k[\phi] \Gamma^{(2)}_k[\phi] k^{-2}) \,.
\eeq    
This is the standard form of the proper-time flow equation.
However, it suffers from unphysical dependencies due to off-shell terms in the rhs.

Here we will amend this equation by an extra term. To motivate this term, we consider allowing the field variable that couples to the source to depend on the cutoff $k$. This leads to 
\beq \label{Gamma}
e^{- \Gamma_k[\phi] } = \int d \hat{\chi}  \, e^{- S[\hat{\chi}]  + ( \hat{\phi}_k[\hat{\chi}] - \phi) \cdot   \frac{ \delta \Gamma[\phi]}{\delta \phi}      }\;.
\eeq
Defining the {\it RG kernel} by
\beq\label{rgkernel}
\Psi_k[\phi]:= \langle  k \partial_k \hat{\phi}_k[\hat{\chi}] \rangle \;,
\eeq
we have that 
\beq \label{Gamma_flow_Psi}
 k \partial_k    \Gamma_k[\phi]  = -  \Psi_k[\phi]  \cdot  \frac{\delta}{\delta \phi}  \Gamma_k[\phi] \,.
\eeq
Adding the two terms of the rhs of \eq{Proper-time} and \eq{Gamma_flow_Psi} we get a {\it generalised proper-time flow equation}
\beq
 k \partial_k    \Gamma_k[\phi]  = -  \Psi_k[\phi]  \cdot  \frac{\delta}{\delta \phi}  \Gamma_k[\phi] + \STr \, \exp(- K^{-1}_k \Gamma^{(2)}_k[\phi]\ k^{-2}) \,.
\eeq
This allows us to implement a perturbative version of the essential RG with a proper-time cutoff. The extra term is proportional to the equations of motion so it lets us remove off-shell terms as we have done by hand in the previous section. Therefore only the essential couplings are compelled to run since we can absorb off-shell terms into $ \Psi_k[\phi]$.

\subsection{The MES to order curvature squared}
Now we apply the essential RG with a proper-time cutoff to order curvature squared neglecting here orders in curvature. We can remove the $R^2$ and $R_{\mu\nu} R^{\mu\nu}$ terms by field redefinitions. Thus, neglecting terms with six derivatives and higher, the action $\Gamma_k$ is given by  
\begin{equation}
\Gamma_k = \int \text{d}^d x \; \sqrt{g} \left(\frac{\rho_k}{8\pi}-\frac{R}{16\pi G_k} +  \vartheta_k \mathfrak{E}(g)  \right) \,.
\end{equation}
Taking into account the ghosts and exploiting the background field method, the generalised proper-time flow equation for gravity takes the form
\begin{equation}
\left( k \partial_k + \int \text{d}^d x  \Psi_{\mu\nu} \frac{\delta}{\delta g_{\mu\nu} } \right)  \Gamma= \Tr \, e^{- K^{-1}  (\Gamma^{(2)}+ S^{(2)}_{\rm gf})k^{-2}}- 2 \Tr \, e^{-\mathcal{Q}_{\rm FP}[\phi]k^{-2}}\;,
\end{equation}
where 
\begin{equation}
	\Psi_{\mu \nu}^g[g] =\gamma_gg_{\mu \nu} + \gamma_R Rg_{\mu \nu} + \gamma_{Ricci }R_{\mu \nu} \;,
\end{equation}
and the equations of motion (Einstein equations) are
\begin{equation}
\frac{\delta \Gamma}{\delta g_{\mu \nu}} = \frac{\sqrt{g}}{2} \frac{\rho}{8\pi} g^{\mu\nu} +\frac{\sqrt{g}}{16\pi G}\left(R^{\mu\nu}-\frac{1}{2}Rg^{\mu \nu}\right)\,.
\end{equation}
 Note that the RG kernel contains all (covariant) terms with up to two derivatives of the metric. This allows us to remove the inessential operators in the action with up to four derivatives. 
As in the last section, we fix $\tilde{\rho}$ to the finite value at the GFP.
This is obtained by going to flat space and setting $\Psi_{\mu\nu} =0$ and looking for the fixed point for $\tilde{\rho} = k^{-d} \rho_k$ (note that now $k$ plays the role of the RG scale $\mu$).
One finds that 
 \beq \label{rho_GFP}
  \tilde{\rho}_{\rm GFP}= (d-3) ( 4 \pi )^{1-\frac{d}{2}} \,.
 \eeq
 Therefore for the proper-time regularisation, we have, from \eq{eta_def} that the essential coupling $\eta$ is 
 \beq \label{etaGtilde}
 \eta =  \tilde{G} ( 4 \pi )^{-\frac{d-2}{2}}\;,
 \eeq
 where now $\tilde{G} = k^{d-2} G_k$.
This differs from the result obtained in dimensional regularisation \eq{eta_MES_DR}.
 
We then obtain the beta functions and gamma functions, with the condition \eq{rho_GFP}, by expanding the trace in the flow equation to order curvature squared.

\subsection{Beta functions and gamma functions}
\begin{figure}
		\centering
		\includegraphics[scale=0.5]{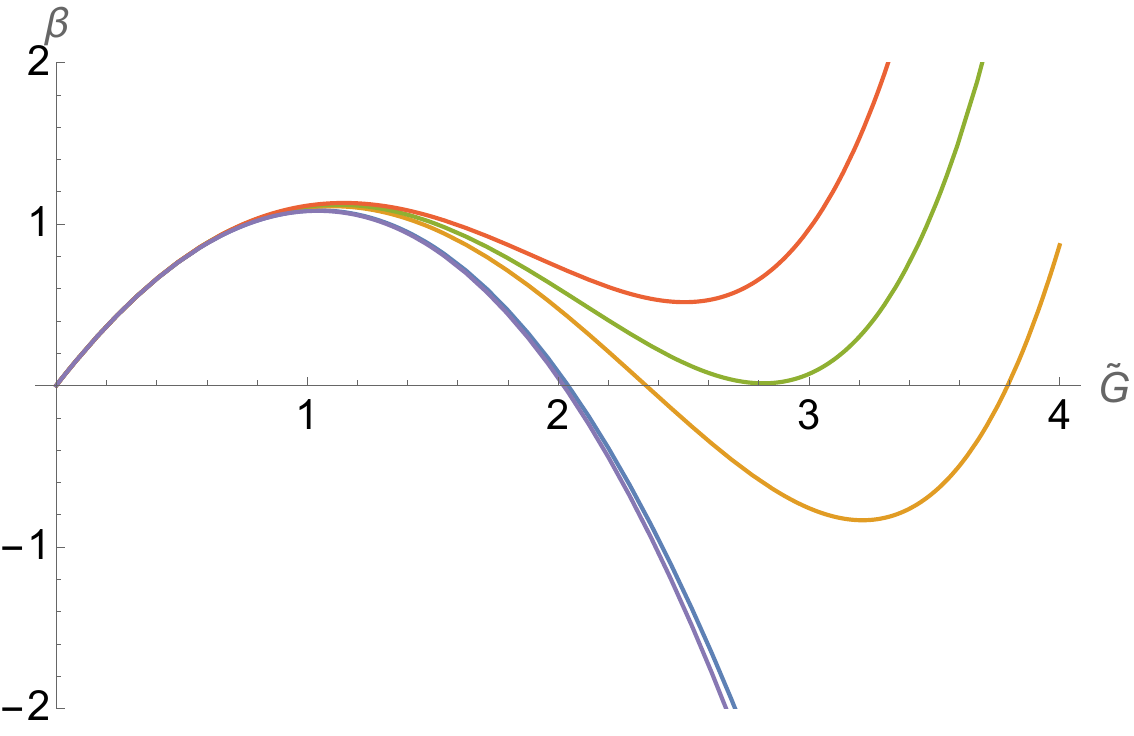}\quad 	
		\caption{The beta function for Newton's constant $\beta(g)$ for different parameterisations. The top curve (red) corresponds to $\tau_1 = -0.8$ which does not feature a fixed point. The second highest curve (green) is for $\tau_1 =-0.705$: in this case two fixed points collide. The third curve (orange) corresponds to $\tau_1 =-0.6$ and the two lowest curves are for $\tau_1 =0$ (blue) and $\tau_1=1$ (purple).   }\label{fig:betatau}
	\end{figure}

We can solve the flow equation for $\gamma_g$, $\gamma_R$, $\gamma_{Ricci}$, $k\partial_k G_k$ and $\partial_k \theta_k$. 
We find
\begin{equation}
    \gamma_g = -2+\frac{2^{3 - d}
   e^{-2 \tilde{G} \rho (-1 + \tau_1 + d  \tau_3)} (-2 + d + d^2 - 
    4 d e^{2 \tilde{G} \rho (-1 + \tau_1 + d  \tau_3} + 
    2 e^{\frac{2 d \tilde{G}\rho(\tau_1 + 2 \tau_2 + 
        d (\tau_3 + 2 \tau_4))}{-2 + d}}) \pi^{1 - d/2}}{d \;\rho}\;.
\end{equation}
The equation for $\beta_{\tilde{G}}$ is a lengthy-expression as a function of the $\tau$'s, $\tilde{G}$ and $d$ (see Fig. \ref{fig:betatau}).
However, if we expand this expression up to order $O(\tilde{G}^3)$ we find
\begin{equation} \label{betag_3}
\beta_{\tilde{G}} = (d-2)\tilde{G} +\frac{1}{3}(-36+(d-3)d)( 4 \pi)^{1-d/2} \tilde{G}^2+ \frac{(d-3)(1200 + d(-566+d(d+19)))  (4\pi)^{2-d}}{30(d-2)}  \tilde{G}^3 + O(\tilde{G}^4)\;.
\end{equation}
Up to this order the beta function does not depend on the parametrisation. Furthermore, if we consider the beta function for $\eta$ by \eq{etaGtilde} we reproduce precisely the beta function \eq{betaeta} obtained in dimensional regularisation. In fact the two schemes are related by sending 
\beq
\tilde{G}_{(\rm PT)} \to  (4 \pi)^{d/2 -1} \tilde{G}_{\rm (DR)}, \,\,\,\,\,\, \tilde{\rho}_{\rm (PT)} \to  (4 \pi)^{-d/2} \tilde{\rho}_{\rm(DR)}\;,
\eeq
where the subscripts indicate the two schemes. We emphasize at this point that this matching is observed by expanding the proper-time beta function up to third order in $G$.
Furthermore, let us stress that both schemes are RG improvements of one-loop RG equations.

Similarly, if we expand $k \partial_k \vartheta_k$ to zeroth order in $G_k$ we obtain
\beq \label{beta_vartheta}
k \partial_k \vartheta_k = k^{d-4} \frac{1}{45} 2^{-d-3} \left(d^2-33 d+540\right) \pi ^{-d/2} + O(G_k)\;,
\eeq
which is independent of parameterisation and is the same as that obtained using dimensional regularisation  \eq{Euler_running_DR} after sending $\theta_k \to (4 \pi)^{d/2 -2} \theta_\mu$ (and $k\to \mu$).

Thus, truncating the beta function to order $\tilde{G}^3$ (or $\eta^3$) we reproduce the results of the previous section and in particular the fixed point and critical exponent \eq{theta_order_2}.
If we do not truncate the beta function for Newton's constant at order $\tilde{G}^3$, it depends on the parameterisation. One then finds that the existence of a fixed point depends on the values of the parameters $\tau_i$. This indicates that the approximation (i.e. second order in curvature and all orders in $G$) is not trustworthy since we can reach different conclusions for different parameterisations of the metric. As an example of the dependence on the $\tau$ parameters, we plot the beta function for Newton's constant in $d=4$ dimensions for different values of $\tau_1$ setting the $\tau_2=\tau_2 = \tau_3=0$ in Fig.~\ref{fig:betatau}.
This shows the danger in coming to any conclusion based on a single parameterisation if this dependence is not under control.
We will not investigate the dependence on the parameterisation further at this order. Instead, we pursue higher order approximations  in curvature where the dependence is absent.

 \subsection{Consistent higher order approximations}
 As we will see in the next section, by going to higher orders in curvature, one can lift the dependence on the parameterisation at higher order in $\tilde{G}$. Hence, at infinite order, there is no dependence on the parametrisation at all. This can be understood since the equation of motion involves terms linear in curvature and terms linear in $G_k \rho_k = G_k  k^d \tilde{\rho}_{\rm GFP}$. Expanding in $G_k$ to all orders and in curvature to a fixed order we will only see the terms proportional to $G_k \rho_k$ in the equation of motion and not the corresponding curvature term, beyond the fixed order. Therefore to correctly resolve terms of the form $\int \text{d}^d x \sqrt{g} R G_k^{N-1}$ on the rhs of the flow equation involves going to order $N$ in curvature. Since on the lhs of the flow equation the beta function appears in the form $ \int \text{d}^d x  \sqrt{g} \beta_{\tilde{G}}/\tilde{G}^2$ going to order $N$ in curvature will resolve $\beta_{\tilde{G}}$ to order $\tilde{G}^{N+1}$. If we look at the beta function for $\tilde{G}$ at higher orders in $\tilde{G}$ then we get a dependence on the parameters $\tau_i$.
 Thus, as long as we work to finite order $N$ in curvature we can approximate the beta function for $\tilde{G}$ by its expansion to order $N+1$. Similarly, the running $\vartheta_k$ at order $N$ in curvature can be approximated by its expansion to order $N-1$ in $\tilde{G}$ to obtain a parameterization independent expression. 

The set of consistent approximations is equivalent to truncating the heat kernel expansion at a given order, where the entire hessian is the operator entering the definition of the heat kernel.
Such an approximation has been put forward in \cite{Falls:2014zba} for essentially the same reasons.

\section{ All curvature orders on a maximally symmetric background}\label{sec:allcurvature}
We now evaluate the proper-time flow equation on a constant curvature maximally symmetric background which allows us to work at all orders in the scalar curvature $R$ and therefore obtain an approximate beta function for Newton's constant that is independent of the parameterization to all orders in the coupling. This is a variant of an ``$f(R)$ approximation'' \cite{Machado:2007ea,Morris:2022btf,Codello:2007bd} for studying the RG flow of quantum gravity but with the advantage that the function $f(R)$ is effectively replaced by a function $\gamma(R)$ which appears in the RG kernel. In particular, we now write the RG kernel introduced in \eqref{rgkernel} as
\beq
\Psi_{\mu\nu} =  \gamma(R) g_{\mu\nu}\;,
\eeq
 which is the most general form on a maximally symmetric background. 
For now, we will neglect the topological term in the action such that $\vartheta_k=0$. In this case the flow equation can be written in the form

\beq \label{flow_cc_background}
\begin{aligned}
\int \text{d}^d x  \sqrt{g} &\left[  \frac{ k^d \beta_{\tilde{\rho}}}{8 \pi } +  \frac{ k^{d-2} \beta_{\tilde{G}} R }{16 \pi \tilde{G}^2 }  - k^{d-2} \frac{(d-2) (\gamma (R)+2) \left(R-\frac{2 d G_k \rho_k }{d-2}\right)}{32 \pi  \tilde{G}} \right] =\\ \qquad&\qquad=  \Tr_0[  e^{-\frac{\Delta_0}{k^2}}]  + \Tr_{2T}[  e^{-\frac{\Delta_2}{k^2}}]    -2 \Tr_1[  e^{-\frac{\Delta_1}{k^2}}] \;,
\end{aligned}
\eeq
with
\bea
\Delta_0 &=& -\nabla^2 -\frac{2 R}{d} +  \sigma
   \left(R-\frac{2 d G_k \rho_k  }{d-2}\right)\\ \label{eq:ghost}
  \Delta_1 &=&  -\nabla^2  - \frac{R}{d}              \\
  \Delta_2 &=&  -\nabla^2 + \frac{2 R}{(d-1) d} + \tau
   \left(R-\frac{2 d G_k \rho_k  }{d-2}\right)\;,
\eea
and the parameters
\bea
\sigma&=&\left(\frac{2 \left(\tau _1-1\right)}{d}+2 d \tau _4+2 \tau _2+2 \tau _3+1\right)\,,\label{eq:sigma}\\
\tau &=&-\frac{(d-2) \left(d \tau _3+\tau _1-1\right)}{d}\,\label{eq:tau}.
\eea
We note that the parameterisation dependence is entirely captured by $\sigma$ and $\tau$ on a constant curvature space.
The evaluation of the traces requires information on the heat kernel coefficients at all orders on a sphere  \cite{Kluth:2019vkg}, which summerised in appendix \ref{app:sphere}. Alternatively, the evaluation via spectral sums. The spectral sums on spheres and hyperboloids are displayed in appendix \ref{app:hyperboloid} (see \cite{Benedetti:2014gja,Falls:2016msz} for previous investigations of the renormalisation group on a hyperboloid).

Following the MES we put $\beta_{\tilde{\rho}} = 0$ and set $\tilde{\rho} = \tilde{\rho}_{\rm GFP}$ given by \eq{rho_GFP}. We can solve \eq{flow_cc_background} for all $R$ in the range $-\infty<R<\infty$ to obtain both beta function $\beta_{\tilde{G}}$ and $\gamma(R)$. Assuming $\gamma(R)$ is regular at the point when the equations of motion 
 \beq
R = \frac{2 d}{d-2} G_k \rho_k =  \frac{2 d}{d-2} k^d G \tilde{\rho}_{\rm GFP}  
 \eeq
are satisfied, we see that we can obtain $\beta_{\tilde{G}}$ independently of $\gamma(R)$ which is multiplied by the equations of motion. The dependence on the $\tau_i$ is also proportional to the equations of motion so  $\beta_{\tilde{G}}$ is independent of the parameterisation to all orders in $\tilde{G}$. For $d>3$ since $R>0$ on the equations of motion we can evaluate the traces on a $d$-sphere\footnote{Technically we truncate the sum over eigenvalues on the sphere to sufficiently high orders which captures the trace for large enough curvature $R/k^2$  \cite{Raul}.} to obtain the beta function for $\tilde{G}>0$.

In $d=4$ we find a fixed point at $\tilde{G} = 2.07$ where the critical exponent is
\beq
\theta = 2.25
\eeq
which is within two percent of the curvature squared result \eq{theta_order_2}. 
It is worth stressing that, while in other approximations that retain all orders in $\tilde{G}$ from a finite order truncation in curvature the fixed point's existence depends on the parameters $\tau_i$, to all orders in $R$ the fixed point always exists and is independent of the $\tau$s.  

\subsection{Inclusion of the topological term}
 The combination $\mathfrak{E} = R^2 - 4 R_{\mu \nu} R^{\mu \nu} + R_{\rho \sigma \mu \nu} R^{\rho \sigma \mu \nu}$ reduces to a curvature squared term on a constant curvature spacetime
 \beq
\mathfrak{E} \to \frac{(d-3) (d-2) R^2}{(d-1) d}\,.
 \eeq
On the other hand for $d=4$ we know that $\int \text{d}^4x \sqrt{g} \, \mathfrak{E}$ is a topological invariant and hence adding a term $\vartheta_k\,  \int \text{d}^4x \mathfrak{E}$ to the action does not affect the hessian and hence does not contribute to the trace. Therefore we can write the action as 
\beq
\Gamma_k = \frac{1}{16 \pi G_k } \int \text{d}^4x \sqrt{g} \left[( 2 \Lambda_k - R)   + \vartheta_k\,  \mathfrak{E} \right]\;.
\eeq
 If we only use information extracted by evaluating the flow equation on a geometry of constant curvature we can not distinguish between $R^2$ terms and the topological term. Therefore we have a term 
 \beq \label{GB_term_cc}
 k \partial_k \vartheta_k\, \int \text{d}^4x \sqrt{g} \mathfrak{E} 
 \to k\partial_k \vartheta_k \, \int \text{d}^4x \sqrt{g}  \frac{(d-3) (d-2) R^2}{(d-1) d}\;,
 \eeq
 which is ambiguous unless we determine $k \partial_k \vartheta_k$ on a more general background. Our previous results in this section assumed $k \partial_k \vartheta_k =0$. However, on the general background, we obtained  \eq{beta_vartheta} from our truncation at order curvature squared. Thus to incorporate the topological term we use \eq{beta_vartheta} neglecting higher orders in $G_k$ since these depend on the parameterisation. This modifies the beta function for Newton's constant since we add \eq{GB_term_cc} to the lhs of \eq{flow_cc_background}. In this case the beta function has a fixed point at $\tilde{G} =2.00$ and a critical exponent 
 \beq \label{theta_all_orders}
\theta = 2.311\;,
 \eeq
which agrees to within half a percent of the curvature squared result \eq{theta_order_2}.
Thus we see that the curvature expansion is very stable as we between $N=2$ and $N = \infty$.
We can also evaluate the traces using the heat kernel expansion keeping terms up to $R^N$ and then expanding the beta function for $\tilde{G}$ to order $\tilde{G}^{N+1}$ (using \eqref{etaGtilde} this can equivalently be identified with the beta function of $\eta$ via the \eq{eta_def} and the renormalization condition \eqref{rho_GFP}). At order $N=5$ and for $d=4$ the beta function is
\bea\label{eq:beta5}
\beta_{\tilde{G}} &=&2 \tilde{G}-\frac{8 \tilde{G}^2}{3 \pi }-\frac{29 \tilde{G}^3}{40 \pi ^2}-\frac{2459
   \tilde{G}^4}{68040 \pi ^3}+\frac{5441 \tilde{G}^5}{3265920 \pi ^4}+\frac{39059
   \tilde{G}^6}{53887680 \pi ^5}+O\left(\tilde{G}^7\right)\,.
\eea
At each order from $N=1$ to $N=5$ we find a fixed point and the critical exponent converges rapidly to the value \eq{theta_all_orders} as can be seen in table~\ref{convergence}. At order $R$ we obtain simply $\theta =2$ since the beta function is expanded to order $\tilde{G}$. At order $R^2$ the beta function is given by \eq{betag_3}.

\begin{table}
\begin{tabular}{c|c}
  ~~~~~~$O(R^N)$~~~~~~&    ~~~~~~$\theta$~~~~~~ \\ \hline
  $R$ & $2$  \\
  $R^2$ & $2.296$   \\
  $R^3 $ & $2.312$ \\
  $R^4 $ & $2.312$\\
  $R^5 $ & $2.311$\\
  $R^\infty $ & $2.311$ 
\end{tabular}
\caption{\label{convergence}Critical exponent at every order of the $R^N$ expansion.   }
\end{table} 

\subsection{The RG kernel}

Having found the fixed point by going on-shell we can then solve for $\gamma_\star(\tilde{R})= \gamma(k^2 \tilde{R})|_{\tilde{G}= \tilde{G}_\star}$, where $\tilde R$ is the dimensionless scalar curvature. To do so we numerically compute the traces of the $4$-sphere and the $4$-hyperboloid. The result is unphysical and depends on the values of the $\tau_i$.  Nonetheless, let us make some remarks about the form of $\gamma_\star(\tilde{R})$. 

\begin{figure}
		\centering
		\includegraphics[scale=0.5]{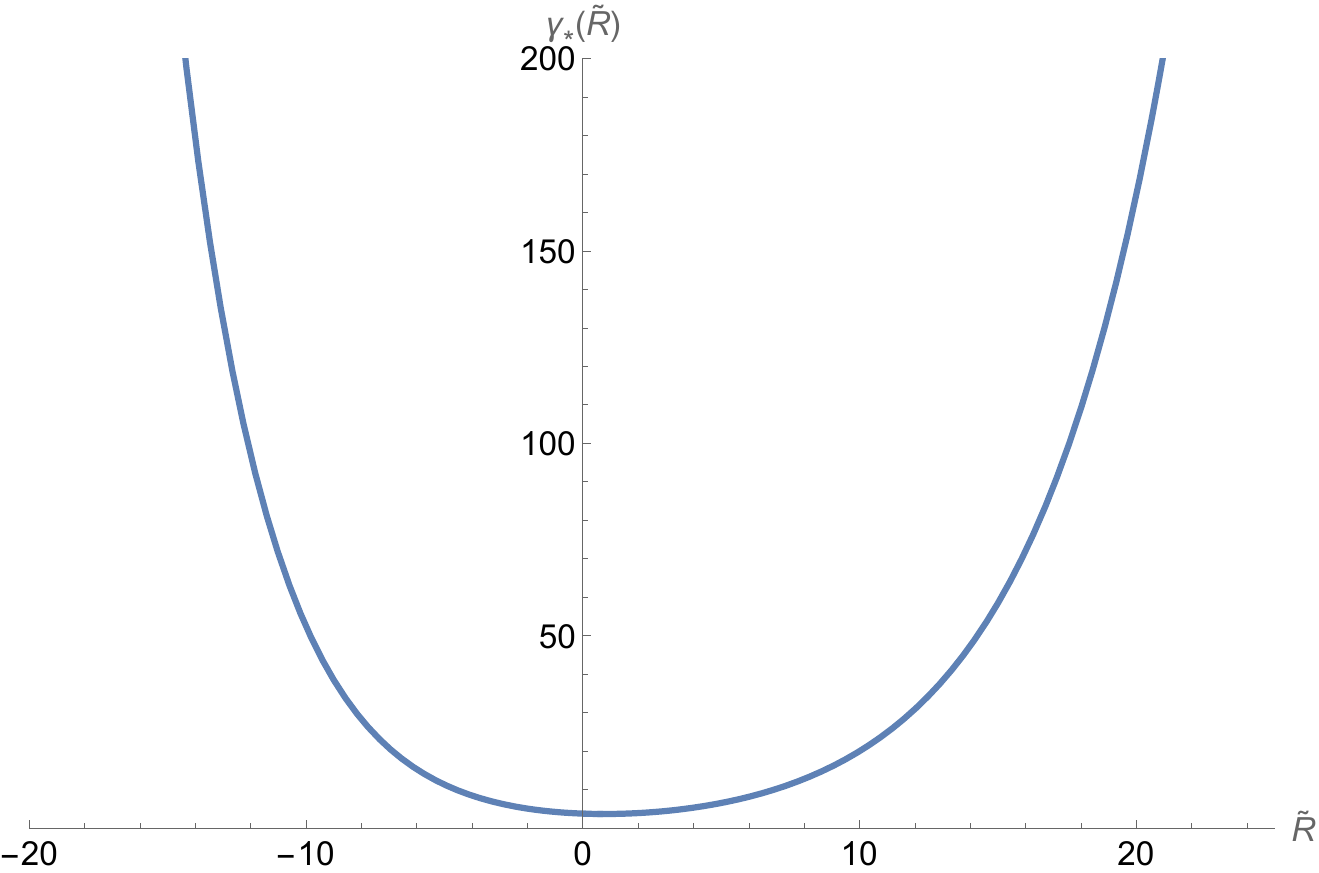}\caption{Plot of $\gamma_\star(\tilde R)$ as a function of the dimensionless Ricci scalar for both hyperbolic ($\tilde R<0$) and spherical  ($\tilde R>0$)  spacetimes with $\tau_i =0$.}\label{fig:gammaR}
	\end{figure}

\begin{figure}
		\centering
		\includegraphics[scale=0.43]{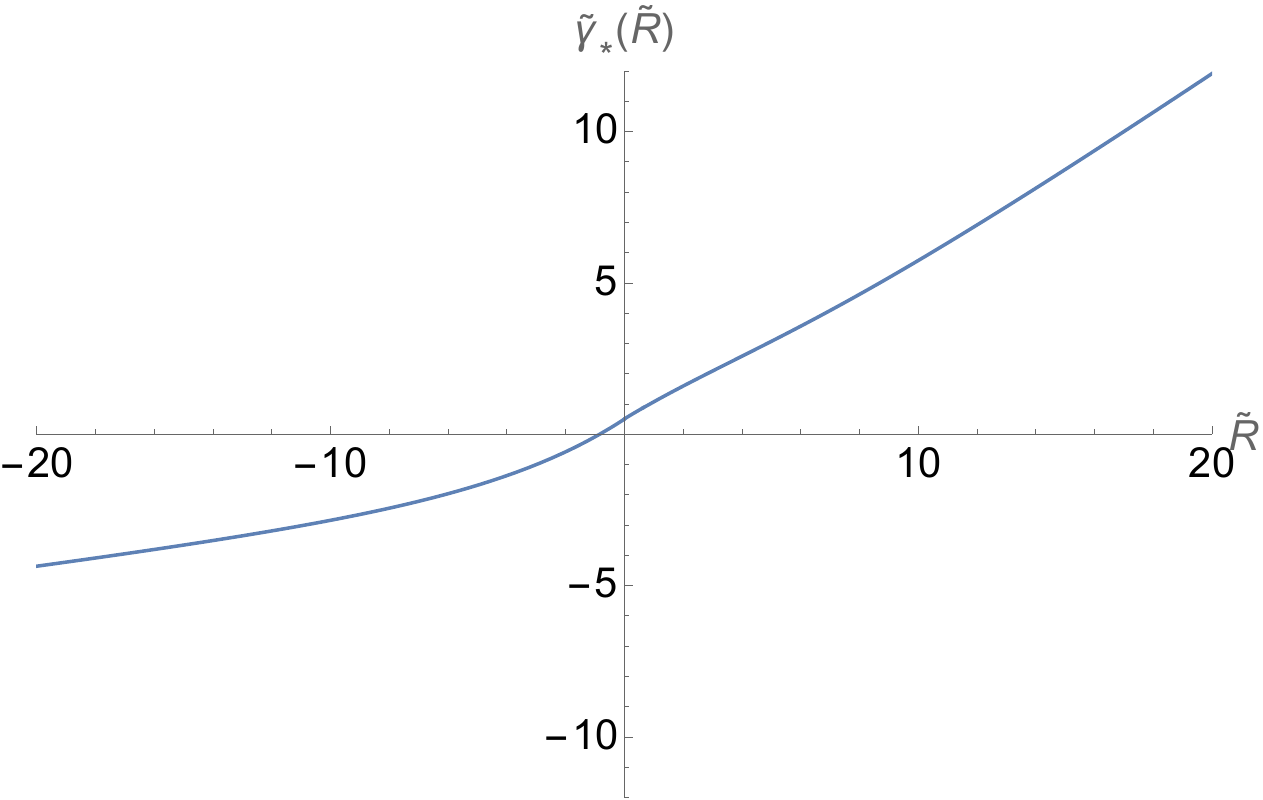}\caption{Plot of $\gamma_\star(\tilde R)$ for  hyperbolic ($\tilde R<0$) and spherical  ($\tilde R>0$)  spacetimes with $\tau =0$ and $\sigma = 10/16$.}\label{fig:gammaR2}
	\end{figure}

As an example, we plot  $\tilde{\gamma}_\star(\tilde{R})$ as a function of the dimensionless scalar curvature $\tilde R$ for $\tau_i=0$ (Fig. \ref{fig:gammaR}). A noticeable feature is that $\tilde{\gamma}_\star(\tilde{R})$ grows exponentially as $\tilde{R}$ grows to large negative or positive values. One can trace this to the fact that there are negative modes for the ghost operator and the hessian (multiplied by the inverse DeWitt metric). For the hessian, the presence of negative modes depends on the values of $\tau_i$ parameters and we can choose them such that all modes are positive. In particular, for $d=4$ we avoid negative scalar modes if the parameter \eqref{eq:sigma} satisfies
\beq
\frac{1}{2} \leq \sigma \leq \frac{11}{16}\,.
\eeq
 The lower bound comes from the sphere and the upper bound from the hyperboloid. Similarly, for the traceless tensor modes, we avoid negative modes when the parameter \eqref{eq:tau} has values in the range
 \beq
-\frac{1}{3} \leq \tau \leq \frac{3}{16}\;.
 \eeq
Technically the Gaussian integral which determines the one-loop effective action does not converge when the parameters are outside this range (implicitly, we are rotating the conformal modes to avoid the instability; we are sending $\bar{g}^{\mu\nu}\hat{h}_{\mu\nu} = \hat{h} \to {\rm i}\hat{h}$ to ensure the convergence of the Gaussian integral over these modes).  The ghost operator, i.e. $\Delta_1$ in \eqref{eq:ghost}, has negative modes on the sphere for our choice of gauge \eqref{eq:gauge}. These correspond to the conformal Killing vectors on the four-sphere.   We can avoid this by choosing a different gauge or rotating this single mode such that the eigenvalue is positive. Indeed one should have the absolute value of the ghost determinant in the functional measure.  
In Fig.~\ref{fig:gammaR2} we have chosen $\tau = 0$ and $\sigma = 10/16$ and we have rotated the negative mode of $\Delta_1$ to demonstrate how $\tilde{\gamma}_\star(\tilde{R})$ behaves without negative modes. We then see that for large curvature the $\tilde{\gamma}_\star(\tilde{R})$ behaves linearly. This happens since the traces become constant for large $\tilde{R}$ while $\tilde{\gamma}_\star$ appears  in the form $\tilde{\gamma}_\star/\tilde{R}$ (note that the volume $\int {\rm d}^d x\sqrt{g}$ scales as $1/R^2$).

We can interpret $\tilde{\gamma}_\star(\tilde{R})$ a defining a field redefinition. Let's set $g_{\mu\nu} =  \omega_k \hat{g}_{\mu\nu}$ with $\omega_k =  \pm 1/R_k$ and  $\hat{g}_{\mu\nu}$ a metric with curvature $\pm 1$ (with the sign given by the sign of $R$).  Then we note that \eq{Gamma_flow_Psi} suggests that $k\partial_k g_{\mu\nu} = \Psi_{\mu\nu}$. From there we obtain
\beq \label{flow_omega}
k \partial_k \tilde{\omega}_k  = 2 \tilde{\omega}_k  +  \tilde{\omega}_k \tilde{\gamma}_\star(\pm 1/\tilde{\omega}_k)\;,
\eeq
where $\tilde{\omega}_k = k^2 \omega_k$. When $\tilde{\gamma}_\star$ behaves as in Fig.~\ref{fig:gammaR} the rhs of \eq{flow_omega} is not a globally Lipschitz since for $\tilde{\omega}_k =0$ the function diverges. We therefore do not expect a global solution for all $k$.  On the other hand, the behaviour seen in  Fig.~\ref{fig:gammaR2} means the rhs of \eq{flow_omega} is globally Lipschitz. We conclude that negative modes should be eliminated to consistently use field redefinitions even if they do not affect the physical results obtained on-shell.

\section{Equivalence between proper-time and dimensional regularisation}\label{sec:equivalence}

Although in the previous section we used the proper-time regulator, here we show that the results will be the same using our variant of dimensional regularisation, at least if we work with the early time heat kernel  expansion.

Let us write a general action in the form
\beq
S = \sum_i \lambda_i \mathcal{O}_i  + S_{\rm ct}\;,
\eeq
where $\lambda_i$ are the renormalised couplings and $O_i$ are different invariants constructed out of the fields, which appear in the early time heat kernel expansion. 
Then the flow equation in  dimensional regularisation can be written in the general form
\beq \label{DR_expansion}
-\mu \partial_\mu  \lambda_i \mathcal{O}_i  =   \sum_{d_c} \frac{\mu^{d-d_c}}{(4\pi)^{d_c/2}} f_{i,d_c}(\lambda,d) \mathcal{O}_i \,,
\eeq
where $f_{i,d_c}(\lambda,d)$ are functions of the couplings and the dimension $d$. If we were doing standard dimensional regularisation we would also replace $ f_{i,d_c}(\lambda,d)\to  f_{i,d_c}(\lambda,d_c)$ and only retain a single term in the sum over $d_c$ i.e. for our chosen $d_c$ e.g. $d_c =4$.

For the proper-time flow with $\Gamma_k = \lambda_i \mathcal{O}_i$, expanding the heat kernel in the early time expansion we will obtain a similar expression
\beq \label{propertime_expansion}
-k \partial_k  \lambda_i \mathcal{O}_i  =  \frac{1}{(4\pi)^{d/2}}  \sum_{d_c} k^{d-d_c} f_{i,d_c}(\lambda,d) \mathcal{O}_i  \;,
\eeq
where the difference is just factors of $4 \pi$ in each term of the beta functions (and $\mu$ is replaced by $k$). If $\mathcal{O}_i$ has dimension $[\mathcal{O}_i] = d_i$ then $f_{i,d_c}(\lambda,d)$ has dimension  $[f_{i,d_c}(\lambda,d)] = - d_i +d_c -d$. Then if in \eqref{propertime_expansion}  we then send
\beq
\lambda_i \to (4\pi)^{-d_i/2} \lambda_i \implies f_{i,d_c} \to (4\pi)^{(d_i-d_c+d)/2}  f_{i,d_c}
\eeq
and compare with \eq{DR_expansion} we see that both equations match (with $\mu$ also replaced by $k$). As a result, the beta functions will only differ by a constant rescaling of the couplings which keeps physical quantities, e.g. scaling exponents, invariant. Note that the rescaling of the couplings keeps dimensionless couplings invariant and can be achieved by a rescaling of the metric tensor $g_{\mu\nu} \to (4 \pi)^{-1} g_{\mu\nu}$.

\section{Critical exponent}\label{sec:crit_exp}
In this section we wish to examine how the exponent $\theta$ defined by 
\beq
\theta  = -\left. \frac{\partial \beta_\eta}{\partial \eta}\right|_{\eta = \eta_\star}  =   -\left. \frac{\partial \beta_{\tilde{G}}}{\partial \tilde{G}}\right|_{\tilde{G} = \tilde{G}_\star} 
\eeq
appears in relations between various scales, namely the bare scale cutoff scale and the observed (i.e. renormalised) Newton's constant or Planck length. In particular, as the bare coupling is sent to the fixed point the cutoff scale in physical units must diverge in a manner determined by $\theta$.

First we consider the running coupling $\tilde{G}(k)$ which goes to zero at the GFP as
\beq \label{g_IR}
\tilde{G}(k\to 0) \to  k^{d-2} G\,,
\eeq
for $d>2$ where $G^{1/(d-2)} \equiv G_{k=0}^{1/(d-2)}$ is the observed Planck length $\ell_P$ . Now if we set the initial condition of $\tilde{G}(k)$ at fixed bare scale $\Lambda$, where $\tilde{G}$ takes the value  $\tilde{G}(\Lambda)$,   then as we send $\tilde{G}(\Lambda) \to \tilde{G}_\star$ the exponent $A$ is defined by
\beq \label{G_UV}
\lim_{\tilde{G}(\Lambda) \to \tilde{G}_\star} G \propto \frac{\Lambda^{2-d}}{|\tilde{G}_\star- \tilde{G}(\Lambda)|^A}\,.
\eeq
For positive $A$ this ensures that, as the bare coupling $\tilde{G}(\Lambda)$ tends to the fixed point value, the physical product of the physical Planck length $\ell_{P} = G^{1/(d-2)}$ and the bare scale $\Lambda$ diverges. In other words, considering
\beq
\xi = \Lambda \ell_{P}\,
\eeq
 as the ``correlation length'' in units of the cutoff, then
\beq
\lim_{\tilde{G}(\Lambda)  \to \tilde{G}_\star} \xi \propto \frac{1}{|\tilde{G}_\star- \tilde{G}(\Lambda)|^\frac{A}{d-2}}\;.
\eeq
To relate $\theta$ and $A$ we note that we can integrate the flow $k \partial_k \tilde{G}(k)  = \beta_{\tilde{G}}$ to obtain
\beq
\log(\Lambda/k) =  \int_{\tilde{G}(k)}^{\tilde{G}(\Lambda)} \frac{1}{\beta_{\tilde{G}}} d\;\tilde{G} \,.
\eeq
Then we note that $\beta_{\tilde{G}}$ has two zeros when we run on a trajectory from the interacting Reuter fixed point to the GFP, at $ \tilde{G}=\tilde{G}_\star$ and $\tilde{G}=0$ respectively.  Supposing we are on such a trajectory, we can write
\beq
\frac{1}{\beta_{\tilde{G}}} = \frac{1}{(d-2) \tilde{G} } -  \frac{1}{\theta (\tilde{G}- \tilde{G}_\star) } + \left(\frac{1}{\beta_{\tilde{G}}} - \frac{1}{(d-2) \tilde{G} } +  \frac{1}{\theta (\tilde{G}- \tilde{G}_\star) }\right) 
\eeq
where the function between the brackets is finite for all $\tilde{G}$ on the trajectory. 
 As such we can write
\beq
\log(\Lambda/k) =  \int_{\tilde{G}(k)}^{\tilde{G}(\Lambda)} \frac{1}{(d-2) \tilde{G} } d\;\tilde{G}  - \int_{\tilde{G}(k)}^{\tilde{G}(\Lambda)} \frac{1}{\theta (\tilde{G}- \tilde{G}_\star) } d\; \tilde{G} + {\rm finite \, terms\, } \,.
\eeq
Integrating this equation in the limit $k \to 0$ and $\tilde G_0 \to \tilde G_\star$ we have
\beq
\Lambda/k \propto  \lim_{k \to 0 , \tilde{G}(\Lambda) \to \tilde{G}_\star}  \left( \tilde{G}(\Lambda)/\tilde{G}(k)\right)^{1/(d-2)} \left( \frac{\tilde{G}(k) - \tilde{G}_\star}{\tilde{G}(\Lambda) - \tilde{G}_\star} \right)^{1/\theta} \;,
\eeq
then using  \eq{g_IR} we obtain  with \eq{G_UV}
\beq
A = \frac{d-2}{\theta}\, .
\eeq

Let us now consider a lattice theory with a lattice spacing $a$ and a dimensionless bare coupling $\kappa$ (see \cite{Hamber:1999nu,Hamber:2009mt} for a general discussion of the methods of quantum gravity on a lattice and the computation of scaling exponents). For different values of the bare coupling, we will produce effective actions with an observed Planck length $\ell_P$. Let us now suppose that we can send $a^{d-2}/G \to0$ by sending $\kappa \to \kappa_\star$ where $\kappa_\star$ is the critical value of the lattice coupling. Then identifying $a = \Lambda^{-1}$ and assuming a non-singular relation between $\tilde{G}$ and $\kappa$ we might expect that the inverse lattice spacing diverges as 
\beq \label{Lattice_theta}
G a^{-d+2}  \propto  \frac{1}{|\kappa_\star- \kappa|^{(d-2)/\theta}}\,.
\eeq

\subsection{Connection with lattice gravity (CDT)}
Given a recent paper \cite{Ambjorn:2024qoe}, we can compare the exponent $\theta$ obtained here, and in FRG calculations, with results coming from lattice simulations using the definition \eq{Lattice_theta} for $d=4$.\footnote{We refer the reader to \cite{Reuter:2011ah, Laiho:2016nlp} for previous attempts to make a connection via the analysis of the spectral dimension.}
From the results of lattice simulations one obtains an effective mini-superspace action of the form \cite{Hartle:1983ai}
\beq \label{Seff}
S_{\rm eff} = \int_{- L \pi \omega /2 }^{L \pi\omega/2 } dt \,  \frac{1}{a^2 \Gamma} \left[  \frac{v'(t)^2}{v(t)} + \delta v^{1/3}(t)  -  \frac{2 \delta ^{3/4}}{L^2}   v(t)       \right]\;, 
\eeq
where $t$ is a time variable, $v(t)$ describes the $3$-volume and the prime denotes a derivative.  Two couplings $\delta$ and $\Gamma$ are present which are determined from the lattice data with
\beq
\omega/\omega_0 = (\delta_0/\delta)^{3/8} \,,
\eeq
where $\omega_0$ and $\delta_0$ are constants.
On the other hand, $L$ is related to the cosmological constant which is fixed in lattice units. In particular $L^4 a^{-4} = \langle N_4 \rangle$ is the expectation value of the number of four simplices, and is fixed by tuning the bare cosmological constant.

The relevant solution to the equations of motion is
\beq \label{vt}
v(t) =L^3 \frac{3}{4 \omega} \cos^3(t/(\omega L ))\,,
\eeq
which satisfies the Euler-Lagrange equations obtained from \eq{Seff}.
Integrating the four-volume is given by 
\beq
 V_4= \int_{- L \pi \omega /2 }^{L \pi\omega/2 } dt v(t) =L^4 \,.
\eeq

Here we want to identify the effective action with the Einstein-Hilbert action obtained in a continuum approach when all fluctuations are integrated out i.e. at $k=0$. There are two issues when making this comparison.
Firstly, when obtained from the Einstein-Hilbert action $\delta$
takes a fixed value $\delta = \delta_0$ and thus the solution \eq{vt} only describes a round four sphere for $\omega=\omega_0$. However, the lattice simulation does not achieve this value. Secondly, in our continuum computations, the cosmological constant is zero at $k=0$ which implies $L^{-1} =0$.

To bring the action into the Einstein-Hilbert form we can make a field redefintion of the form
\beq
v(t) = z V(\tau)\,,\,\,\,\,\,\,    \tau = z t\;,
\eeq
which means that
\beq
V_4 =  \int dt z V(zt) = \int d\tau V(\tau)\;.
\eeq
In other words, we have rescaled space and time while keeping the four-volume invariant.  Then the effective action is
\beq
S_{\rm eff} = \int_{- L z \pi \omega /2 }^{L \pi z \omega/2 } d\tau \,  \frac{z^2 }{a^2 \Gamma} \left[   \frac{V'(\tau)^2}{V(\tau)} + \delta z^{-8/3} V^{1/3}(\tau)   - \frac{2 \delta ^{3/4}}{L^2} z^{-2}   V(\tau) \right] \,,
\eeq
and the solution is given by
\beq
V(\tau) = z^{-1} L^3 \frac{3}{4 \omega} \cos^3(\tau/( z \omega L ))\,.
\eeq
We can then choose
\beq
z = (\delta/\delta_0)^{3/8} = \omega_0/\omega
\eeq
such that the effective action is of the Einstein-Hilbert form
\beq
S_{\rm eff} = \int d\tau \,  \frac{\omega_0^2 }{ \omega^2 a^2 \Gamma} \left[   \frac{V'(\tau)^2}{V(\tau)} + \delta_0 V^{1/3}(\tau)   -  \frac{2 \delta_0 ^{3/4}}{L^2}   V(\tau)  \right] \,,
\eeq
and the solution to the equations of motion is 
\beq
V(\tau) =  L^3 \frac{3}{4 \omega_0} \cos^3(\tau/(  \omega_0 L ))\;,
\eeq
which describes a four sphere of volume $L^4$.\footnote{We refer the reader to  \cite{Ferrero:2022hor} for a similar discussion related to self-consistent spacetimes, namely that background chosen for a specific RG scale which satisfies the equations of motion at that scale.}
Then we can identify $G$ and $\rho$ by   
\beq
 G  =  (\omega/\omega_0)^{2} \frac{\Gamma}{24 \pi} a^2\,,\,\,\,\,\,\,\,\,  G \rho =  \frac{2 \delta_0 ^{3/4}}{L^2}\,,
\eeq
and the dimensionless coupling is given by
\beq
4\pi \eta^2 \equiv G^2 \rho =    (\omega/\omega_0)^{2} \frac{\Gamma}{24 \pi} a^2 \frac{2 \delta_0 ^{3/4}}{L^2} = (\omega/\omega_0)^{2} \frac{\Gamma}{24 \pi}  \frac{2 \delta_0 ^{3/4}}{\sqrt{ \langle N_4 \rangle}}\,.
\eeq
Since $G^2 \rho$ is a physical dimensionless coupling the lattice RG flow takes it to be fixed. In each individual simulation $\langle N_4 \rangle$ and the limit $\langle N_4 \rangle \to \infty$ must be extrapolated.   

 As a putative critical point is approached we can define the two critical exponents $\alpha$ and $\beta$ by
\beq
\Gamma \propto  \frac{1}{|\kappa_\star- \kappa|^\alpha}\,, \,\,\,\,\,\,\,\,\,\,\,\, \,\, \omega \propto  |\kappa_\star- \kappa|^\beta\,\,\,\, \implies \,\,\,\, G a^{-2} \propto  \frac{1}{|\kappa_\star- \kappa|^{\alpha-2 \beta}}
\eeq
 and hence identify $A = \alpha -2 \beta$ and the exponent $\theta$ as
 \begin{equation}
     \theta_\text{CDT}= \frac{2}{\alpha-2 \beta}\,.
 \end{equation}
 A similar reasoning has lead to the same relation \cite{Ambjorn:2024qoe}.
According to the lattice simulations performed there, the value of this exponent is 
\beq
\theta_\text{CDT} = 4 \pm 1\;,
\eeq
which is almost twice the size of our exponent. 
Hence, from this result one could conclude that our model belongs to a different universality class than the one of CDT. However, for more definitive conclusions,  a clearer connection between the two approaches and a more controlled comparison among the respective assumptions and approximations would be needed.

\section{Conclusion}\label{sec:conclusions}
Due to diffeomorphism invariance, it is a longstanding open question, how to derive observable quantities in quantum gravity. Typically a gauge fixing procedure is needed to single out physical degrees of freedom.
Furthermore, even before fixing the gauge, there is much freedom in gravity to parameterise the geometric degrees of freedom.
However, in concrete computations, this choice gets entangled along the different steps, and it is a difficult task to trace it back, eventually, to remove the dependencies on this choice afterward. The most natural choice is to use the equations of motion to derive on-shell quantities such as the S-matrix.

On another level, standard perturbative methods fail to be predictive for some theories, resulting in a non-renormalisability of the associated operators. In particular, Einstein gravity is non-renormalisable in dimensional regularisation at two-loops. Non-perturbative methods have however shed light on this, by a UV completion realised via a non-trivial fixed point, rendering the theory predictive. Using those procedures, however, it can be difficult to disentangle physical quantities from spurious effects due to necessary approximations. 

In this paper we used a variant of dimensional regularisation that generalises the scheme of \cite{Martini:2021slj} where divergences that appear as  $d \to 2$ were investigated and the equations of motion were used to remove terms that depend on unphysical choices. Keeping poles that are present in {\it all dimensions}, we have a regularisation that keeps all power-law and logarithmic divergences in four dimensions. At one-loop this scheme gives a renormalisation group flow that can be mapped to the proper-time equation by multiplying the couplings by a constant factor.

At this stage, it is important to emphasize the similarities and differences with the recent publication of \cite{Kluth:2024lar}. There the non-minimal subtraction scheme of \cite{Martini:2021slj} was not used and only poles in two and four dimensions played a role. Here we have used the non-minimal subtraction scheme which allowed for the map to the proper-time to be established. Consequently, this lets us work fully functionally in $R$ such that poles in all even dimensions are subtracted.

 Following the spirit of the essential RG \cite{Baldazzi:2021orb}, we derived the generalised proper-time flow.
Here we adopted the MES, identifying inessential couplings close to the GFP and then constraining the form of the effective action such that the inessential couplings are zero around the GFP. In particular, in gravity, the vacuum energy  $\rho$ was identified as the inessential coupling to be fixed to its value at the GFP \cite{Baldazzi:2021orb}.\footnote{Interestingly, research on the $N$-cutoffs has found evidence pointing in a similar direction from a  different point of view \cite{Ferrero:2024yvw}. Here, also, on-shellness is imposed on the solution in a self-consistent manner.}

In regards to the critique of asymptotic safety and the meaning of running couplings \cite{Donoghue:2019clr}, we note that our approach concentrates on the essential couplings that enter observables. General arguments dictate that if these couplings approach a UV fixed point observables will be free from unphysical divergencies \cite{Weinberg:1976xy,Weinberg:1980gg,Baldazzi:2021orb}. Our approach, based on dimensional regularisation, demonstrates that such a fixed point can be found in perturbation theory and without the use of a cutoff function.

To test our scheme,   we employed a general ultra-local parameterisation of the  metric in terms of a fluctuation field. In general, the untruncated beta function for Newton’s constant $\tilde{G}=k^{d-2} G_k$ will depend on the different parameterisations.   Terms which depend on the choice of the parameterisation through the parameters $\tau_i$ are proportional to the equation of motion. On the other hand, the freedom to make field redefinitions also produces terms proportional to the equations of motion on the lhs of our flow equation. By consistently truncating the expansion in curvature and in Newton's constant we have seen that the essential beta functions are independent of the parameters $\tau_i$. We expect that the beta function will also be gauge-independent, but it remains to show this explicitly.   

Despite the very different approaches to the RG our result for the critical exponent $\theta$ computed agrees well with the value found using the NPRG with the MES \cite{Baldazzi:2021orb,Knorr:2022ilz,Knorr:2023usb,Baldazzi:2023pep}. 
For example, the value of $\theta$ in the range $2.1 \pm 0.2$ appeared favored in \cite{Baldazzi:2023pep} after scanning a wide range of regulators.
In that work the flow equations were solved including all terms with six derivatives of the metric. In \cite{Knorr:2023usb} the full momentum dependence at order curvature squared was taken into account and the value $\theta = 2.347$ was obtained.

Finally, we compared our result for the critical exponent with the value obtained in a recent publication in CDT \cite{Ambjorn:2024qoe}. We considered the scaling relations of the lattice putative fixed point, which we can relate to the critical exponent of the essential coupling $\eta$. The two results are not yet in good agreement. 
One possibility is that CDT and the model of quantum gravity considered here, based on the quantisation of the metric field, are not in the same universality class and hence do not share the same scaling exponents.

\bigskip
The present work lays the ground for further studies regarding the application of the scheme to different aspects of gravity. First of all, it would be critical to perform this investigation at two-loop order. Standard perturbative methods lead to the conclusion that gravity is a non-renormalisable theory at two-loop level. This computation will involve the treatment of two-loop diagrams and higher order poles and will represent an important test for the methods presented here. Furthermore, we plan to further investigate the relation with the scaling exponents computed in lattice gravity. It would be particularly important to test the exponents computed along the second order phase transition.
As an important generalisation, this treatment can be extended to Lorentzian spacetimes. Recently, several techniques to solve the flow in Lorentzian signature have been developed (see for instance \cite{Manrique:2011jc,Biemans:2016rvp,Fehre:2021eob,Banerjee:2022xvi,DAngelo:2022vsh,DAngelo:2023wje, Saueressig:2023tfy,Thiemann:2024vjx, Ferrero:2024rvi}) and would find also an application in the essential RG, with the aim to compute real-world observables (see \cite{Baldazzi:2021fye} for a discussion on relational observables in asympototic safety).

\section*{Acknowledgements}
We would like to thank Jan Ambjørn,  Alessandro Codello, Yannick Kluth, Riccardo Martini, Martin Reuter, Marc Schiffer and Omar Zanusso for useful discussions and comments. RF is grateful for the hospitality of Perimeter Institute where part of this work was carried out. Research at Perimeter
Institute is supported in part by the Government of Canada through the Department of Innovation, Science and
Economic Development and by the Province of Ontario through the Ministry of Colleges and Universities. This work
was supported by a grant from the Simons Foundation (1034867, Dittrich).

\appendix

\section{One-loop traces for the linear parameterization}
\label{app_trace}
In this appendix we give a pedagogical derivation of the one-loop traces for Einstein gravity in the linear parametrization $g_{\mu\nu} = \bar{g}_{\mu\nu} + h_{\mu\nu}$ i.e. $\tau_i =0$. In particular we  detail the proper time or heat kernel representation of the one-loop traces.

Defining the Green's function $G$ as the inverse of the operator in \eqref{1-loop}
\beq
G^{-1} =  K^{-1} (S^{(2)} + S^{(2)}_{\rm gf}) \,,
\eeq
we can express 
\beq
G= \frac{1}{\Delta+ U + E}\,,
\eeq
where $\Delta = - \nabla^2 \mathbb{1}$ is the Laplacian times the identity for symmetric tensors
\beq
 \mathbb{1}_{\mu\nu}\,^{\rho\lambda}(x,y) = \frac{1}{2}( \delta^{\rho}_{\mu}\delta^{\lambda}_{\nu} + \delta^{\lambda}_{\mu}\delta^{\rho}_{\nu}) \delta(x-y)\,,
\eeq
and the potential-like terms using the linear parametrization are a term with zeroth order in curvature
\begin{equation}
    (U)_{\mu\nu}\,^{\rho\lambda} = -2\rho G   \mathbb{1}_{\mu\nu}\,^{\rho\lambda} 
\end{equation}
and a term linear in curvature
\bea
(E)^{\mu\nu,\rho \sigma}(x,y) &=& -\frac{1}{d-2}\left(R g^{\mu \nu} g^{\rho \sigma} -(d-2) R^{ \mu \nu}g^{\rho \sigma}-(d-2)(R^{\mu \rho}g^{\sigma \nu}+R^{\nu \rho}g^{\sigma \mu})+2 R^{\rho \sigma}g^{\mu \nu})\right)\nonumber\\&&\qquad\qquad -(-R g^{\mu \nu} g^{\rho \sigma} +(R^{\mu\rho\nu\sigma}+R^{\mu\sigma\nu\rho})) \delta(x-y)
\eea
Note that throughout the appendix we have already identified the full metric with the background metric $\bar g = g$ and hence used $g$ to lower and raise indexes.

It is useful to define 
 \beq
 G_0= \int_0^\infty \text{d}s e^{-s (\Delta + U)} \;. \label{G_0}
 \eeq
Then, we can expand \eqref{1-loop} to second order in $E$, i.e., in curvature, by the expansion 
\begin{equation}
	\begin{aligned}
\log \left[ K^{-1} (S^{(2)} + S^{(2)}_{\rm gf})\right] =&- \log\left(G\right)
\\
=&-\log\left(G_0\right) +EG_0-\frac{EG_0EG_0}{2}+...
	\end{aligned}
\end{equation}
Hence, using \eqref{G_0}, we can write
\begin{equation}
	\begin{aligned}
		\log\left(G_0\right) =\log\int_0^\infty \text{d}s e^{-s (\Delta + U)}  = \int_0^\infty \frac{\text{d}s}{s} e^{-s (\Delta + U)}
	\end{aligned}
\end{equation}
and also
\begin{equation}
	\begin{aligned}
		EG_0 =E\int_0^\infty \text{d}s e^{-s (\Delta + U)}  \,,
	\end{aligned}
\end{equation}
and
\begin{equation}
	\begin{aligned}
		EG_0EG_0 =E^2\int_0^\infty \text{d}s s e^{-s (\Delta + U)} \,, 
	\end{aligned}
\end{equation}
where we neglect covariant derivatives of the curvature since these lead to boundary terms which we neglect. 
We can now trace by exploiting the heat kernel expansion up to order curvature square (see appendix \ref{app:off})
\begin{equation}
\begin{aligned}
\text{Tr}\log\left(K^{-1}S_\text{grav}^{(2)}\right) &= -\text{Tr}\log\left(G_0\right) +\text{Tr}EG_0-\text{Tr}\frac{EG_0EG_0}{2}\\& = -\int \frac{\text{d}s}{s}	\Tr [e^{-s(\Delta+U)}]	+\int ds	\Tr [e^{-s(\Delta+U)}E]- \frac{1}{2}	\int \text{d}ss	\Tr [e^{-s(\Delta+U)}	E^2]\\&=-\int \text{d}s\frac{1}{(4\pi s)^{d/2+1}}\int \text{d}^d x \sqrt{g} \left\{\tr \mathbb{1}+\frac{s}{6}R \tr \mathbb{1}\right.\\
&\qquad\left.+s^2\left(\frac{1}{180}\left(R_{\mu \nu \rho \sigma}R^{\mu \nu \rho \sigma} + R_{\mu \nu}R^{\mu \nu} + D^2\; R\right)\;\tr \mathbb{1} \right.\right. \\&\qquad\left.\left.+ \frac{1}{72}R^2\;(\tr\mathbb{1})^2 + \frac{1}{12}[\nabla_\mu, \nabla_\nu][\nabla^\mu, \nabla^\nu] \tr\mathbb{1}+ \frac{1}{36} \Delta R\; \tr\mathbb{1}\right)\right\}
\\&\quad+\int \text{d}s\frac{1}{(4\pi s)^{d/2}}\int \text{d}^d x \sqrt{g} \left\{\tr [\mathbb{1}E]+\frac{s}{6}R \tr[ \mathbb{1}E]\right\}\\&\quad-\frac{1}{2}\int \text{d}s\frac{1}{(4\pi s)^{d/2-1}}\int \text{d}^d x \sqrt{g}\left\{\tr [\mathbb{1}E^2]\right\}\,.
\end{aligned}
\end{equation}
To this expression, we  add the terms coming from the ghosts, which are computed similarly.

\section{Evaluation of non-commuting traces}\label{app:noncomm}
 In this appendix we evaluate the same one-loop traces for a general parametrisation as in \eqref{eq:expand}. We emphasise that the dependence on the $\tau$'s and hence on the parametrisation would disappear from the hessian if one imposes the validity of the equations of motion for the metric.
 Tracing the heat kernel expansion for a general parameterisation requires us to take into account the non-commutativity of operators appearing in the hessian. In this appendix we will show a useful strategy to evaluate those non-commuting traces.
 
The hessian in \eqref{Proper-time} can be split in the following contributions
 \beq\label{B1}
K^{-1} (S^{(2)} + S^{(2)}_{\rm gf}) = - \nabla^2 \mathbb{1} + U(\tau_i) + E(\tau_i)
 \eeq
where now $U$ and $E$ depend on the values of the $\tau_i$ parameters and coincidence with the expressions of the previous section when $\tau_i =0$.
 $E$ is the endomorphism which is proportional to curvature. For a general parametrization \eqref{eq:expand} the endomorphism and the coefficients are given by  
 \bea
E_{\mu\nu}\,^{\rho\sigma}(x,y)& =&\Bigg[ \frac{1}{d-2}\Big((-1 +\tau_1+\tau_2+ (d-2)\tau_3+2(d-2)\tau_4)R  g_{\mu \nu}  g^{\rho \sigma} -(d-2)(1+\tau_2) R_{ \mu \nu} g^{\rho \sigma}\nonumber\\
&&-2(d-2)(\tau_1-1)(R^{\rho}\,_\mu \delta^{\sigma}_{ \nu}+R^{\rho}\,_\nu \delta^{\sigma}_\mu)-2(\tau_1+\tau_2-1) R^{\rho \sigma} g_{\mu \nu}\Big)\nonumber\\
&&-\Big((-1+\tau_1+(d-2)\tau_3)R  g_{\mu \nu}  g^{\rho \sigma} +(R_{\mu}\,^\rho\,_\nu\,^\sigma+R_{\mu}\,^\sigma\,_\nu\,^\rho)\Big) \Bigg] \delta(x-y)\,.
\eea
$U$ is given by 
\beq
U = -a \mathbb{1}- b P\,,
\eeq
with
\begin{equation}
a= -2\Lambda (-1 + \tau_2 + d \tau_3)\,,
\end{equation}
\begin{equation}
b=\frac{2 d \Lambda (\tau_1 + 2 \tau_2 + d\tau_3 + 
    2 d\tau_4)}{  d-2}\,,
\end{equation}
and where $P$ is the projector onto the trace 
 \beq
 P_{\mu\nu}\,^{\rho\lambda}(x,y) = \frac{1}{d} g_{\mu\nu} g^{\rho\lambda} \delta(x-y)\,,    
 \eeq
 which satisfies $P^2 = P$.
Everything commutes (up to derivatives of curvature) apart from $P$ and $E$.
This means we can write the trace over the exponential of the hessian as 
 \beq
 \Tr \, e^{-  (- \nabla^2 - a \mathbb{1} + E    - b P  )k^{-2} } =  \Tr \, e^{-  (- \nabla^2 - a \mathbb{1} ) k^{-2} }  e^{-( E    - b P  )k^{-2} }
 \eeq
Then we can expand to second order in $E$ to obtain
\bea \label{eq:App_expansion}
\Tr \, e^{-  (- \nabla^2 - a \mathbb{1} + E    - b P  )k^{-2} } &=& \Tr \, e^{-  (- \nabla^2 - a) k^{-2} }  (1+ f_1(b/k^2) P) \nn
&& - k^{-2} \Tr \, e^{-  (- \nabla^2 - a) k^{-2} }   E   (1+ f_1(b/k^2) P)\nn  &&+ \frac{1}{2}  k^{-4}  \Tr \, e^{-  (- \nabla^2 + a) k^{-2} } \left(  E^2 + f_1(b/k^2) P E^2  + f_2(b/k^2) P [E,P]E ) \right)  \\
&&+ \dots \nonumber
\eea
where the dots represent higher orders in $E$ and terms involving derivatives of curvature, and 
 \beq
 f_1(\tilde{b}) =  e^{\tilde{b}} -1 \;,
 \eeq
 \beq
 f_2(\tilde{b}) = \frac{e^{\tilde{b} } \tilde{b} +\tilde{b} -2 e^{\tilde{b} }+2}{\tilde{b} }.
 \eeq
The expansion \eq{eq:App_expansion} can be realised by using, for example, the identity
\beq
\delta \, e^X = \int_0^1 ds e^{(1-s) X} \delta X e^{s X} 
\eeq
for an operator $X$, and the fact that $P^2 = P$.

\section{Off-diagonal heat kernel}\label{app:off}
In this appendix we report the (off-)diagonal heat kernel expansion \cite{Groh:2011dw,Ferrero:2023xsf} used up to curvature square in section \ref{sec:allcurvature}. The heat kernel expansion reads
\begin{eqnarray}
H &=&\frac{1}{( 4\pi s)^{d/2} }(A_0 +s\,A_1)\,,\\
	H_\mu &=&\frac{1}{( 4\pi s)^{d/2}}(\mathcal{D}_\mu A_0 +s\,\mathcal{D}_\mu \,A_1)\,,\\
H_{(\mu \nu)} (x,s) &=& \frac{1}{( 4\pi s)^{d/2} }\left(-\frac{1}{2s}g_{\mu \nu}A_0-\frac{1}{2}g_{\mu \nu}A_1+\mathcal{D}_{(\mu}\mathcal{D}_{\nu)}A_0 \right)\\
H_{(\mu \nu \rho \sigma)} (x,s) &=& \frac{1}{( 4\pi s)^{d/2} }\left(\frac{3}{4}\frac{1}{s^2} g_{(\mu \nu}g_{\rho \sigma)}A_0 +\frac{3}{4} \frac{1}{s}g_{(\mu \nu}g_{\rho \sigma)}A_1+ \frac{3}{4}g_{(\mu \nu}g_{\rho \sigma)} A_2 \nonumber \right.\\&&\left.-\frac{3}{s}g_{\rho \sigma} \mathcal{D}_{(\mu}\mathcal{D}_{\nu)}A_0-3 g_{\rho \sigma} \mathcal{D}_{(\mu} \mathcal{D}_{\nu)}A_1 +  \mathcal{D}_{(\mu} \mathcal{D}_{\nu}\mathcal{D}_{\rho}\mathcal{D}_{\sigma)}A_0\right)\,,
\end{eqnarray}
where $s$ represents the proper-time. The coefficients read
\begin{eqnarray}
A_0 &=&1 \,, \hspace{2cm}
\mathcal{D}_\mu A_0 =0 \,,\hspace{2cm}
\mathcal{D}_{(\mu}\mathcal{D}_{\nu)} A_0=\frac{1}{6}R_{\mu \nu}\\
\mathcal{D}_{(\mu}\mathcal{D}_{\nu} \mathcal{D}_{\rho}\mathcal{D}_{\sigma)}A_0&=&\frac{3}{10} \nabla_{(\alpha}\nabla_{\beta}R_{\nu \mu)}+ \frac{1}{12}R_{(\alpha \beta}R_{\mu \nu)}+ \frac{1}{15} R_{\gamma(\beta|\delta|\alpha}R^\gamma{}_\nu{}^\delta{}_{\mu)}\\
A_1& =& -E+\frac{1}{6}R \,,\hspace{2cm}
\mathcal{D}_\mu A_1  =  -\frac{1}{2}\nabla_\mu E + \frac{1}{6}\nabla_\nu\Omega{}^\nu{}_{\mu} +\frac{1}{12}\nabla_\mu R\\
\mathcal{D}_{(\mu}\mathcal{D}_{\nu)} A_1&=&-\frac{1}{3}\nabla_{(\mu}\nabla_{\nu)}E-\frac{1}{6}R_{\mu \nu} E-\frac{1}{6}\nabla^\alpha \nabla_{(\nu} \Omega_{\alpha) \mu}+ \frac{1}{6}\Omega_{\alpha(\nu}\Omega^\alpha{}_{\mu)}+\frac{1}{20}\nabla_{(\mu}\nabla_{\nu)}R\nonumber\\
&&-\frac{1}{60}\Delta R_{\mu \nu} + \frac{1}{36}RR_{\mu \nu} -\frac{1}{45}R_{\nu \alpha}R^{\alpha}{}_\mu + \frac{1}{90}R_{\alpha \beta}R^{\alpha}_\nu{}^\beta{}_\mu+ \frac{1}{90} R^{\alpha \beta \gamma}{}_\nu R_{\alpha \beta \gamma \mu}\\
A_2 &=& \frac{1}{6}\Delta E + \frac{1}{2}E^2 - \frac{1}{6}R E + \frac{1}{12}\Omega_{\mu \nu}\Omega^{\mu \nu} - \frac{1}{30}\Delta R \nonumber\\&& + \frac{1}{72}R^2 -\frac{1}{180}R_{\mu \nu}R^{\mu \nu}+\frac{1}{180}R_{\mu \nu \alpha \beta}R^{\mu \nu \alpha \beta}\;,
\end{eqnarray}
where $\Omega_{\mu \nu} = [\nabla_\mu, \nabla_\nu]$. These coefficients will be used to evaluate the traces in section \ref{sec:dim_reg}.

\section{Heat kernel expansion on a sphere}\label{app:sphere}
In this appendix we will summarize the results of \cite{Kluth:2019vkg}, which we used to evaluate the all curvature order expansion of the trace.

The heat kernel $H(s,x,y)$ is defined as the solution to the equation
\begin{equation}
    \frac{\partial H(s,x,y)}{\partial s} = (\nabla^2+E)H(s,x,y)\;,
\end{equation}
where $E$ is an endomorphism.
The formal solution is given by
\begin{equation}
H(s,x,y)= e^{s(\nabla^2+E)}
\end{equation}
For early times there exist an expansion of  the heat kernel as an asymptotic series following the DeWitt ansatz
\begin{equation}
H(s,x,y)= \frac{\Delta^{1/2}}{(4 \pi s)^{d/2} } \text{exp}\left\{\frac{-\sigma}{2s}\right\}\sum_{n = 0}^{\infty} \left[\tilde b_{2 n} (E, x,y)s^n+ \tilde c_{d+2n}(E, x,y) s^{d/2+n}\right]\;,
\end{equation}
where $\Delta$ is the Van Fleck-Morette determinant and $\sigma$ is the Synge's world function.  The coefficients $\tilde{c}_{d+2n}$, which are not part of the standard DeWitt ansatz, appear when we consider constrained fields such as a transverse vector \cite{Kluth:2019vkg}.

Since we are ultimately interested in the trace of the heat kernel, we only need the coincidence limit of $H$
\begin{equation}
    \text{Tr} \;[H(s)] = \frac{1}{(4 \pi s)^{d/2} } \sum_{n = 0}^{\infty} \left[\text{Tr}\;[\tilde b_{2 n} (E)]s^n+ \text{Tr}\;[\tilde c_{d+2n}(E) ]s^{d/2+n}\right]\;.
\end{equation}
For a compact spacetime one can define
\begin{equation}
b_{2 n} (E) = \frac{1}{\text{Vol}}\;\; \text{Tr}\;[\tilde b_{2 n} (E)]\; \qquad\;c_{d+2 n} (E) =\frac{1}{\text{Vol}}\;\; \text{Tr}\;[\tilde c_{d+2n} (E)]\;,
\end{equation}
which allows us to write
\begin{equation}
 \text{Tr}\;[ H(s)] = \frac{\text{Vol}}{(4 \pi s)^{d/2} } \sum_{n = 0}^{\infty} \left[\text{Tr}\;[b_{2 n} (E)]s^n+ \text{Tr}\;[c_{d+2n}(E) ]s^{d/2+n}\right]\;.
\end{equation}
For maximally symmetric spacetimes, the volume of a $d$-dimensional sphere reads
\begin{equation}
    \text{Vol} = \frac{2 \pi^{(d+1)/2}}{\Gamma\left(\frac{d+1}{2}\right)} \left(\frac{d(d-1)}{R}\right)^{d/2}\;.
\end{equation}

Importantly, one can relate the heat kernel having a given endomorphism $E$ ($H_E$) by that with another endomorphism $\bar E$ ($H_{\bar E}$) via
\begin{equation}
H_E(s, x,y) = e^{-s(E- \bar E)} H_{\bar {E}} (s, x, y)
\end{equation}
implying that the  coefficients are related through
\begin{equation}
b_{2n} (E) = \sum_{k = 0}^n \frac{(E-\bar E)^k}{k!} b_{2(n-k)}(\bar E)
\end{equation}
\begin{equation}
c_{d+2n}(E) = \sum_{k = 0}^n \frac{(E-\bar E)^k}{k!} c_{d+2(n-k)}(\bar E)
\end{equation}

In ref. \cite{Kluth:2019vkg} the scalar, vector and tensor coefficients have been computed explicitly. As an  example, for scalar fields (denoted by the over-script $(0)$ from the 0-spin) one has
\begin{equation}
    b_{2n}^{(0)} = \frac{\Gamma(1-d/2)}{\Gamma(1-n-d/2)} \left(\frac{R}{d(d-1)}\right)^n \kappa_n(d)\;,
\end{equation}
\begin{equation}
    c_{d+2n}^{(0)} = 0\;,
\end{equation}
where $\kappa_n(d)$ is determined by the generating function
\begin{equation}
 \text{exp} \; \sum_{n=1}^\infty \frac{(-1)^{n+1}}{n(2n+1)}B_{2n+1}\left(\frac{d-1}{2}\right)z^n = \sum_{n = 0}^\infty\kappa_n(d)z^n
\end{equation}
and $B_n(\cdot)$ is the $n$-th Bernoulli polynomial.

\section{Spectral sum on spheres and hyperboloids}\label{app:hyperboloid}
In this appendix we will report the spectral sums method used in section \ref{sec:allcurvature}. We follow the conventions in \cite{Benedetti:2014gja, Falls:2016msz,Banerjee:2023ztr}.

The spectrum of the Laplacian  on the $d$-sphere for fields of arbitrary spin $j$ is well known:
\begin{equation}
-\nabla^2 \phi_{n,l} =\omega_n\phi_{n,l} \;,
\end{equation}
where  $\omega_n$ is the $n$-th eigenvalue
\begin{equation}
    \omega_n = \frac{R}{d(d-1)}(n(n+d-1)-j)\;, \qquad n \in \mathbb{N}
\end{equation}
with $D_n$  degeneracy number
\begin{equation}
    D_n^{(0)} = \frac{(n+d-2)!(2n+d-1)}{n!(d-1)!}\;,\qquad n = 0,1, \dots\;,
\end{equation}
\begin{equation}
    D_n^{(1)} = \frac{n(n+d-1)(n+d-3)!(2n+d-1)}{(n+1)!(d-2)!}\;,\qquad n = 1,2, \dots\;,
\end{equation}
\begin{equation}
    D_n^{(2)} = \frac{(d+1)(d-2)(n+d)(n-1)(n+d-3)!(2n+d-1)}{2(n+1)!(d-1)!}\;,\qquad n = 2,3, \dots\;,
\end{equation}
for scalars, vectors and tensors, respectively.

 For the sphere we can then use the spectral sum representation of the heat trace, which is convergent in the large-$s$ domain,
 \begin{equation}
\text{Tr}\;[e^{\nabla^2s}]_{(S^d)} = \frac{1}{\text{Vol}}  \sum_n D_n e^{-s\omega_n} = \frac{\Gamma\left(\frac{d+2}{2}\right)}{2 \pi^{(d+1)/2}} \left(\frac{R}{d(d-1)}\right)^{d/2} \sum_n D_n e^{-\omega_ns} \;.
\end{equation}
Finally the trace over an arbitrary function $W$ of the Laplacian (including an endomorphism) reads:
 \begin{equation}
\text{Tr}\;[W(-\nabla^2+E)]_{(S^d)} = \frac{\Gamma\left(\frac{d+2}{2}\right)}{2 \pi^{(d+1)/2}} \left(\frac{R}{d(d-1)}\right)^{d/2} \sum_n D_n W(\omega_n+E)\;.
\end{equation}

For the hyperboloid, using standard polar coordinates for the line element on the sphere $\text{d}s^2 = a^2 \text{d} \xi^2 +a^2 \sin^2(\xi)\text{d}\Omega_{d-1}^2$, the scalar eigenfunctions and associated eigenvalues on the hyperboloid can
be obtained through the analytic continuation
\begin{equation}
    \xi = i y\;, \qquad n = -\rho + i \lambda\;, \qquad \rho = \frac{d-1}{2}\;.
\end{equation}
This results in the practical relation between the eigenvalues of the sphere and those of the hyperboloid:
\begin{equation}
\big(\mathcal{E}_\lambda\big){}_{\mathbb{H}(a)} = -\big(\mathcal{E}_n\big){}_{\mathbb{S}(a)}\Big|_{n = -\rho +  i \lambda}\;, \qquad \lambda \in \mathbb{R}_{\geq0}
\end{equation}
Explicitly, for $  \mathbb{H}_d$ the scalar spectral problem reads
\begin{equation}
    -\nabla^2\varphi_{\lambda,l} = \nu_\lambda\varphi_{\lambda,l}
\end{equation}
with eigenvalues 
\begin{equation}
\nu_\lambda = -\frac{R}{d(d-1)}(\lambda^2+\rho^2+j)\;, \qquad \lambda \in \mathbb{R}_{\geq0} \end{equation}

Next, we would like to consider fields with arbitrary spin $j$. Here, one has to distinguish between even and odd dimensions. In even dimensions one finds
 \begin{equation}
\text{Tr}\;[e^{\nabla^2 s}]_{(H^d)} = \frac{2^{d-2}g(j)}{\pi}  \frac{\Gamma\left(\frac{d}{2}\right)}{2 \pi^{d/2}} \left(\frac{-R}{d(d-1)}\right)^{d/2}\int_0^\infty \text{d}\lambda \;\mu(\lambda )\; e^{-\nu_\lambda s},
\end{equation}
where we have used the (Plancherel) spectral measure 
 \begin{equation}
\mu(\lambda) \equiv  \frac{2 \pi^{d/2}}{\Gamma\left(\frac{d}{2}\right)} \left(\frac{-R}{d(d-1)}\right)^{-d/2}\frac{\pi}{2^{d-2}g(j)}\sum_{l}\varphi^*_{\lambda,l}(0)\varphi_{\lambda,}(0)= \frac{\pi (\lambda^2+(j + \frac{d-3}{2})^2)\lambda\tanh(\pi \lambda)}{2^{2(d-2)}\Gamma(d/2)^2}\prod_{l = \frac{1}{2}}^{\frac{d-5}{2}} (\lambda^2+l^2)\;,
\end{equation}
with
\begin{equation}
    g(j) = \frac{2j+d-3(j+d-4)!}{(d-3)!j!}\;.
\end{equation}
 Note that for $d= 4$ the product is omitted.
 Finally the trace over an arbitrary function $W$ of the Laplacian (including an endomorphism) reads:
 \begin{equation}
\text{Tr}\;[W(-\nabla^2 +E)]_{(H^d)} = \frac{2^{d-2}g(j)}{\pi} \frac{\Gamma\left(\frac{d}{2}\right)}{2 \pi^{d/2}} \left(\frac{-R}{d(d-1)}\right)^{d/2} \int_0^\infty \text{d}\lambda \;\mu(\lambda )\; W(\nu_\lambda +E)\;.
\end{equation}


\end{document}